\documentstyle[onecolumn,psfig]{mn}

\newif\ifAMStwofonts

\ifoldfss
  \ifCUPmtlplainloaded \else
    \NewTextAlphabet{textbfit} {cmbxti10} {}
    \NewTextAlphabet{textbfss} {cmssbx10} {}
    \NewMathAlphabet{mathbfit} {cmbxti10} {} % for math mode
    \NewMathAlphabet{mathbfss} {cmssbx10} {} %  "   "    "
  \fi
  \ifAMStwofonts
    \ifCUPmtlplainloaded \else
      \NewSymbolFont{upmath} {eurm10}
      \NewSymbolFont{AMSa} {msam10}
      \NewMathSymbol{\upi}     {0}{upmath}{19}
      \NewMathSymbol{\umu}     {0}{upmath}{16}
      \NewMathSymbol{\upartial}{0}{upmath}{40}
      \NewMathSymbol{\leqslant}{3}{AMSa}{36}
      \NewMathSymbol{\geqslant}{3}{AMSa}{3E}

      \let\leq=\leqslant 
      \let\geq=\geqslant 
    \fi
  \fi
\fi % End of OFSS

\ifnfssone
  \newmathalphabet{\mathit}
  \addtoversion{normal}{\mathit}{cmr}{m}{it}
  \addtoversion{bold}{\mathit}{cmr}{bx}{it}
  \newmathalphabet{\mathbfit} % math mode version of \textbfit{..}
  \addtoversion{normal}{\mathbfit}{cmr}{bx}{it}
  \addtoversion{bold}{\mathbfit}{cmr}{bx}{it}
  \newmathalphabet{\mathbfss} % math mode version of \textbfss{..}
  \addtoversion{normal}{\mathbfss}{cmss}{bx}{n}
  \addtoversion{bold}{\mathbfss}{cmss}{bx}{n}
  \ifAMStwofonts
    \ifCUPmtlplainloaded \else
      \UseAMStwoboldmath
      \makeatletter
      \new@mathgroup\upmath@group
      \define@mathgroup\mv@normal\upmath@group{eur}{m}{n}
      \define@mathgroup\mv@bold\upmath@group{eur}{b}{n}
      \edef\UPM{\hexnumber\upmath@group}
      \new@mathgroup\amsa@group
      \define@mathgroup\mv@normal\amsa@group{msa}{m}{n}
      \define@mathgroup\mv@bold\amsa@group{msa}{m}{n}
      \edef\AMSa{\hexnumber\amsa@group}
      \makeatother
      \mathchardef\upi="0\UPM19
      \mathchardef\umu="0\UPM16
      \mathchardef\upartial="0\UPM40
      \mathchardef\leqslant="3\AMSa36
      \mathchardef\geqslant="3\AMSa3E

      \let\leq=\leqslant 
      \let\geq=\geqslant 
    \fi
  \fi
\fi % End of NFSS release 1

\ifnfsstwo
  \DeclareMathAlphabet{\mathbfit}{OT1}{cmr}{bx}{it}
  \SetMathAlphabet\mathbfit{bold}{OT1}{cmr}{bx}{it}
  \DeclareMathAlphabet{\mathbfss}{OT1}{cmss}{bx}{n}
  \SetMathAlphabet\mathbfss{bold}{OT1}{cmss}{bx}{n}
  \ifAMStwofonts
    \ifCUPmtlplainloaded \else
      \DeclareSymbolFont{UPM}{U}{eur}{m}{n}
      \SetSymbolFont{UPM}{bold}{U}{eur}{b}{n}
      \DeclareSymbolFont{AMSa}{U}{msa}{m}{n}
      \DeclareMathSymbol{\upi}{0}{UPM}{"19}
      \DeclareMathSymbol{\umu}{0}{UPM}{"16}
      \DeclareMathSymbol{\upartial}{0}{UPM}{"40}
      \DeclareMathSymbol{\leqslant}{3}{AMSa}{"36}
      \DeclareMathSymbol{\geqslant}{3}{AMSa}{"3E}

      \let\leq=\leqslant 
      \let\geq=\geqslant 
    \fi
  \fi
\fi % End of NFSS release 2

\ifCUPmtlplainloaded \else
  \ifAMStwofonts \else % If no AMS fonts
    \def\upi{\pi}
    \def\umu{\mu}
    \def\upartial{\partial}
  \fi
\fi

\title[Millisecond and Binary Pulsars as Nature's Frequency Standards]
{Millisecond and Binary Pulsars as Nature's Frequency Standards. \\
{\LARGE III. Fourier Analysis and Spectral Sensitivity of Timing Observations 
             to Low-Frequency Noise}}
\author[Sergei M. Kopeikin and Vladimir A. Potapov]
       {Sergei M. Kopeikin $^{1,2}$
       and Vladimir A. Potapov $^2$\\ \\
      $^1$ TPI, FSU Jena,  Max-Wien-Platz 1, D - 07743, Jena, Germany\\
      $^2$ PRAO, ASC FIAN, Leninskii Prospect, 53, Moscow 117924, Russia}
\date{Accepted.............199... ;  
      Received ............199... ; 
      in original form ...........199... }
\pubyear{1999}

\def\LaTeX{L\kern-.36em\raise.3ex\hbox{a}\kern-.15em
    T\kern-.1667em\lower.7ex\hbox{E}\kern-.125emX}

\begin{document}
\Large
\label{firstpage}

\maketitle

\begin{abstract}
Millisecond and binary pulsars are the most stable natural
frequency standards which admits to introduce modified versions of 
universal and
ephemeris time scales based correspondingly on the intrinsic rotation of
pulsar and on its orbital motion around barycenter of a binary system. 
Measured stability of these time scales depends on numerous physical
phenomena which
affect rotational and orbital motion of the pulsar and observer on the 
Earth, perturb 
propagation of electromagnetic pulses from pulsar to the observer and bring 
about random
fluctuations in the rate of atomic clock used as a primary time reference
in timing observations. On the long time
intervals the main reason for the instability of the pulsar time scales is 
the presence of correlated, 
low-frequency timing noise in residuals of times of arrivals (TOA) of pulses
from the pulsar
which has both astrophysical and geophysical
origin. Hence, the timing noise can carry out the important physical
information about interstellar medium, interior structure of the pulsar,
stochastic gravitational waves coming from the early universe, etc. 
Each specific type of the low-frequency noise can be 
described
in framework of power law spectrum model. Although the data processing of
pulsar timing observations in time domain seems to be the most imformative 
it is significantly
important to know to which
spectral bands single and binary pulsars, considered as detectors 
of the low-frequency noise signal, are the most sensitive. Solution of this
problem may be reached only if a parallel processing of timing data in frequency
domain is fulfilled. This requires a development of the Fourier analysis
technique for an adequate interpretation of data contaminated by the correlated
noise with a singular spectrum. 
The given problem is examined in the present article.   
\end{abstract}

\section{Introduction}

Millisecond and binary pulsars are known as exellent probes for testing
theory of general relativity Taylor \& Weisberg 1982, 1989), 
structure of interstellar medium (Rickett 1990,
1996) and interior
of neutron stars (Cordes \& Greenstein 1981, Kaspi {\it et al.} 1994)
as well as setting upper limit on the energy density of 
primordial
gravitational radiation (Kaspi , Thorsett \& Dewey 1996,
McHugh {\it et al.} 1997, Kopeikin 1997a, Kopeikin \& Wex 1999). 
Rotational motion of a pulsar around its own axis had been 
proposed (Shabanova {\it et al.} 1979, Backer {\it et al.} 1982, 
Rawley {\it et al.} 1987, Matsakis {\it et al.} 1997) for using as a 
new time reference being analogue of
universal time in astrometry. Quite recently, a
new step toward to estiblishing a stable time scale on extremely long 
intervals approaching 50-100 years has been suggested 
(Ilyasov {\it et al.} 1998, Kopeikin 1999).
It is extracted from the orbital motion of pulsar in a binary system and
represents the analogue of ephemeris time of classical astronomy introduced
by Newcomb (1898)at the end of last century and based on the orbital motion of
the Earth around the Sun.

An adequate analysis of timing data requires deeper understanding of nature of
a noise process dominating in pulsar timing residuals. As soon as the
autocovariance function of the noise process is known the analysis in time
domain becomes possible. Time domain analysis is the most informative since
the observed stochastic process is not usually stationary and includes 
the non-stationary component as well (Groth 1975, Cordes 1978, 1980; Kopeikin
1997b). Because pulsar timing observations are conducted on relatively
long time intervals the white noise of errors in measuring TOA of pulsar's 
pulses will be suppressed by the presence of a number of 
correlated, low-frequency (red) noises having different spectra and
intencities. Henceforth, we are mainly interested in analysing the
red noise. 

The simple model of such noise has
been worked out by one of us (Kopeikin 1997b). It is based on the shot
noise approximation and do includes a dependence of the autocovariance
function on both stationary and non-stationary components of red noise. In the
process of elaborating the given model a rather
remarkable fact has been established (Kopeikin 1999), namely, that timing 
residuals and variances of some spin-down and all orbital parameters are not
affected by the non-stationary component of the red noise at all. 
This discovery put on a firm ground the  
Fourier analysis of TOA residuals and variances of fitting parameters 
in frequency domain. This analysis gives exhaustive information
about the noise process itself and admits us to reveal to which
frequency harmonics in the spoectral expansion of the stochastic process 
pulsar timing observations are the most sensitive. Moreover, just we have an
adequate approach for the treatment of red noise in frequency domain a lot of
interesting applications is opened having the goal to study the physical 
nature of the low frequency noises with arbitrary spectrum.       

Any low-frequency noise can be approximately 
characterized by the power-law spectrum $S(f) \sim
f^{-n}$ where the spectral (integer) index $n \geq 1$. 
It is obvious that the spectrum has a singularity
at zero frequency. Hence, the energy of TOA residuals comprised in
such noise should has infinite value because the integral over all frequencies
from zero to infinity taken from $S(f)$ is divergent. Clearly, this has no
physical meaning and one has to resort to special mathematical tricks
in order to avoid this artificial divergency. There are two ways for 
curing this flaw. First of
them consists in analytical continuation of the spectrum by changing it from
$S(f)\sim f^{-n}$ to $S(f,A)\sim f^{-(n+A)}$ where $A$ is the pure complex 
parameter being differnt 
from zero. Such model of the analitically continued spectrum gives convergent 
integrals which coincide everywhere
on real axis with ones taken from the spectrum $S(f)\sim f^{-n}$ exept 
for the point $f=0$. In order to
prescribe a physical meaning to such integrals we have to expand them in
the Laurent series with respect to the parameter A and to take the finite part
of the expansion. Such procedure has been used, in particular, 
by Kopeikin (1997a) for
calculation of autocovariance function of stochastic noise of the primordial 
gravitational wave background having spectrum $S(f) \sim f^{-5}$ (Mashhoon
1982, 1985, Mashhoon \& Seitz 1991, Bertotti {\it et al.} 1983). 

The procedure of analytical
continuation of divergent inetgrals is mathematically rigorous and
theoretically powerful tool which gives well-defined and self-consistent
results (Gel'fand \& Shilov 1964). However, it is inconvinuent for 
people doing numerical computations.
For them, the second method of regularization of singular spectra seems 
to be more preferable and practically useful. It is based on the
truncation of all divergent integrals at the lower cut-off frequency
$f=\varepsilon$ along with corresponding modification of the power-law model of
the red noise spectrum in order to avoid the model dependence of results of
fitting procedure on the artificially introduced cut-off frequency. 
It will be shown that the simplest way of modification of the spectrum can be 
reduced to the 
addition to the existing power-law spectrum of red noise of the infinite sum
consisting of the Dirac delta function and its derivatives having local 
support at
the lower cut-off frequency $f=\varepsilon$. Such modernization of the spectrum
preserves the structure of the autocovariance function and,
as a consequence, do not vilolate results of numerical computations 
in time domain. This second method of 
regulariazation of divergent integrals will be used in the present paper.  

In what follows it is more convinuent for us to work in terms of dimensionless
frequency and time. For instance, in binary pulsars it is 
preferable for analytic calculations 
to measure time in units of orbital frequency $n_b=2\pi/P_b$, where
$P_b$ is the orbital period of the binary. Then, frequency $f$ is measured in
units of $1/P_b$ and dimensionless time is the pulsar's mean anomaly
$u=n_b\tau$. Hereafter, we use $u$ instead of $\tau$.

\section{Regularized Spectrum of Low-Frequency Noise}

Any gaussian low-frequency noise is completely characterized by the 
autocovariance
function which describes correlation between two values of stochastic process
separated by the arbitrary time interval $\tau=t_2-t_1$. Autocovariance
function consists usually of two algebraically independent parts characterizing
separately stationary $R^-(\tau)$ and non-stationary $R^+(\tau)$ 
components of the noise. Complete expressions of 
autocovariance functions for
different events of low-frequency noise have been derived in the paper 
(Kopeikin 1997b) where the shot noise approximation of stochastic process has
been used. Although both stationary and non-stationary components of
autocovariance function are important for an adequate treatment of observations
(Kopeikin 1999) we are dealing in the present paper only with the stationary
part which posseses to be transformed into the spectral density of noise
$S(f)$ by means of the Wiener-Khintchine theorem
\vspace{0.3 cm}
\begin{equation}
\label{aa}
R^-(u)=2\int_0^{\infty}S(f)\cos(2\pi f u)df.
\end{equation}
If (constant) intensity of noise is denoted by $h_n$ then 
autocovariance function of
low-frequency correlated noise is determined by the expression (Kopeikin 1997b)
\vspace{0.3 cm}
\begin{equation}
\label{gr}
\left\{\begin{array}{ll}C_n h_n |u|^{n-1}, &\mbox{$n=2,4,6,...$,
random walk noise}\\ \\
C_n h_n u^{n-1} \ln|u|, 
& \mbox{$n=1,3,5,...$, flicker noise}
\end{array}\right. 
\end{equation} 
where $C_n$ is a numerical constant of normalization.   
 
Functions, which
might be appropriate candidates for the spectrum of noise procesess with the
foregoing autocovariance functions, are $S(f)={\rm const}f^{-n}$ where $n$ is
integer. However,
integrals (\ref{aa}) from such power-law functions are divergent because of
non-physical singularity at zero frequency. For this reason, regularization
technique should be used because we don't know usually the low-frequency
behavior of the spectrum. 

\subsection{Analaitic Continuation Technique}

Analytic continuation
regularization procedure is contained in that one extends the spectrum $S(f)$ 
in the
complex plane domain by introducing the function 
\vspace{0.3 cm}
\begin{equation}
\label{ab}
S(f,A)={\rm const}f^{-n-A}
\end{equation}
where $A$ is a complex parameter. Autocovariance function 
becomes an analytically
continued functional of the complex variable $A$:
\vspace{0.3 cm}
\begin{equation}
\label{ac}
R^-(u,A)=2\int_0^{\infty}S(f,A)\cos(2\pi f u)df,
\end{equation}
which coincides (due to the properties of analytically continued complex
functions) exactly with the integral in (\ref{aa}) exept at the point $A=0$.
The functional (\ref{ac}) with the spectrum defined by eq. (\ref{ab}) is a 
table integral and can be easily calculated analitically. After calculation of
the integral it is expanded in the Laurent series near the point $A=0$. It
yields
\vspace{0.3 cm}
\begin{equation}
\label{ad}
R^-(u,A)=\frac{1}{A}{\rm Residue}_{A=0}\left\{R^-(u,A)\right\}
+R^-(u,A=0)+...\quad .
\end{equation}
The first term in the expansion is a simple pole with respect to $A$. The second
term in the expansion is finite and gives exactly the autocovariance function
given in eq. (\ref{gr}). Regularization of the integral (\ref{ac}) means that
we take only its finite part and abandon the singular term. Analytic
continuation technique is powerful theoretical tool (Kopeikin 1997a) 
but it hardly can be used in
numerical computations for which infrared cut-off technique is much better.  

\subsection{Infrared Cut-off Technique}

The power spectrum $S(f)$ of the noise is defined 
using the truncated Fourier transform with the lower cut-off frequency
$f=\varepsilon$.  
Namely, we require that the 
truncated cosine Fourier transform of $S(f)$ must give the stationary part 
of the original autocovariance function (\ref{gr}) without any 
additional contributions. Let us postulate that the spectrum $S(f)$ may be
represented by the formula
\vspace{0.3 cm}
\begin{equation}
S(f)=\left\{\begin{array}{ll}h_n\left[\displaystyle{
\frac{1}{(2\pi f)^n}}+\displaystyle{\sum_{k=0}^\infty}B_{2k}(\varepsilon) 
\varepsilon^{2k}\delta^{(2k)}(f-\varepsilon)\right], &\mbox{if
$f\geq\varepsilon$}\\ \\
0 , & \mbox{otherwise}
\end{array}\right.
\label{1}   
\end{equation}
\vspace{0.3 cm}
where the spectral index of noise $n=1,2,...,6$, constant parameter $h_n$ 
is the strength of noise, quantities 
$B_k(\varepsilon)$ are constant numerical coefficients being defined
later, and $\delta^{(k)}(f-\varepsilon)$ denotes the $n-th$ derivative 
with respect to $f$
of the Dirac delta-function $\delta(f-\varepsilon)$. The Dirac delta function
is defined according to the condition (Korn \& Korn 1968)
\begin{equation}
\displaystyle{\int_a^b f(x)\delta(x-X)dx}=\left\{\begin{array}{ll}0,
&\mbox{if $X<a$, or $X>b$,}\\ \\
\frac{1}{2}f(X+0),&\mbox{if $X=a$,}\\ \\
\frac{1}{2}f(X-0),&\mbox{if $X=b$,}\\ \\
\frac{1}{2}\left[f(X-0)+f(X+0)\right],&\mbox{if $a<X<b$,}
\end{array}\right.
\label{1aa}
\end{equation}
\vspace{0.3 cm}
where $f(x)$ is arbitrary function being such that unilateral limits $f(X-0)$
and $f(X+0)$ exist.
Coefficients $B_k(\varepsilon)$ are determined by the condition that 
the cosine Fourier transform
of $S(f)$ gives stationary part of autocovariance function of corresponding
low-frequency noise the model of which may be found in papers (Kopeikin 1997b,
1999).  
As an example of derivation of $B_k(\varepsilon)$ we determine the
several first
coefficients $B_k(\varepsilon)$ in the event of flicker noise in pulsar's
rotational phase 
which has the spectral
index $n=1$. Coefficients $B_k(\varepsilon)$ for noises having other 
spectral indices will be displayed in this section without proof which is
rather straightforward. 

Stationary part of autocovariance function $R^{-}(\tau)$ of flicker noise in 
pulsar's phase is 
equal to $h_1 \pi^{-1} \log|u|$. 
According to definition of the spectrum we should have
\vspace{0.3 cm}
\begin{equation}
R^{-}(u)=2\displaystyle{\int_{\varepsilon}^{\infty}}S(f) \cos(2\pi f u)df.
\label{2}
\end{equation}
\vspace{0.3 cm}
Substituting $S(f)$ from Eq. (\ref{1}) with $n=1$ 
to right hand side of Eq. (\ref{2}) and taking integrals we get
\vspace{0.3 cm}
\begin{equation}
R^{-}(u)=-\frac{h_1}{\pi}{\bf Ci}(2\pi \varepsilon |u|)+
h_1\cos(2\pi\varepsilon u)
\displaystyle{\sum_{k=0}^\infty}(-1)^k B_{2k}(\varepsilon) 
(2\pi\varepsilon u)^{2k},
\label{3}   
\end{equation}
\vspace{0.3 cm}
where ${\bf Ci}(x)$ is the cosine integral, and 
we have used the formula (Korn \& Korn 1968)
\vspace{0.3 cm}
\begin{equation}
2\displaystyle{\int_{\varepsilon}^{\infty}}\delta^{(2k)}(f-\varepsilon)
\cos(2\pi f u)df=\frac{d^{2k}}{d\varepsilon^{2k}}
\cos(2\pi\varepsilon u)=
(-1)^k (2\pi u)^{2k}\cos(2\pi\varepsilon u)
\label{4}
\end{equation}
\vspace{0.3 cm}
Taylor expansion of the cosine integral and $\cos(2\pi\varepsilon u)$ in the 
right hand side of the expression (\ref{3}) with respect 
to small parameter $\varepsilon$ yields
\vspace{0.3 cm}
\begin{equation}
R^{-}(u)=\frac{h_1}{\pi}\left[-\log |u|-\gamma-\log(2\pi\varepsilon)\right]+
h_1\biggl\{B_0(\varepsilon)+(2\pi\varepsilon u)^2
\left[\frac{1}{2\pi}-\frac{1}{2}B_0(\varepsilon)-
B_2(\varepsilon)\right]\biggr\}+O(\varepsilon^4).
\label{5}
\end{equation}
\vspace{0.3 cm}
Since $R^{-}(u)$ must be equal to $-h_1\pi^{-1}\log |u|$, we find 
from Eq. ({\ref 5})
\vspace{0.3 cm}
\begin{equation}
B_0(\varepsilon)=\frac{\gamma}{\pi}+\frac{c\log(2\pi\varepsilon)}{\pi},\hspace{2 cm}
B_2(\varepsilon)=\frac{1}{4\pi}-\frac{\gamma}{2\pi}-\frac{\log(2\pi\varepsilon)}{2\pi}
\label{6}
\end{equation}
\vspace{0.3 cm}
For practical purposes it is enough to account for the coefficient 
$B_0(\varepsilon)$ only,
since all other residual terms appear being multiplied by the factor 
$2\pi\varepsilon u$
which is negligibly small under the usual circumstances because of smallness of the 
product $\varepsilon u$. Thus, residual terms are not important as long as
the accuracy of observations is not high enough. It is worth emphasizing 
that the residual terms under
discussion are model dependent. Had we chosen another model for the spectrum
having slightly another behavior as frequency approaches to zero 
the residual terms would look differently. These arguments permit to estimate
how long we can observe a certain pulsar and process the data with one or 
another model of red noise spectrum.   

Proceeding in the same way for the set of other spectral indeces we obtain 
the following expressions for the power spectra of low-frequency noises:

\begin{enumerate}
\renewcommand{\theenumi}{(\arabic{enumi})}

\item \underline{Flicker noise in phase}:
\vspace{0.3 cm}
\begin{equation}
S(f)=\frac{h_1}{2\pi}\biggl\{
\frac{1}{ f}+2
\left[\gamma+\log(2\pi\varepsilon)\right]\delta(f-\varepsilon)\biggr\}+
O(\varepsilon^2).
\label{7}   
\end{equation}
\vspace{0.3 cm}

\item \underline{Random walk in phase}:
\vspace{0.3 cm}
\begin{equation}
S(f)=\frac{h_2}{4\pi^2}\biggl\{
\frac{1}{f^2}-
\frac{2}{\varepsilon}\delta(f-\varepsilon)\biggr\}+
O(\varepsilon).
\label{8}
\end{equation}
\vspace{0.3 cm}

\item \underline{Flicker noise in frequency}:
\vspace{0.3 cm}
\begin{equation}
S(f)=\frac{h_3}{8\pi^3}\biggl\{
\frac{1}{f^3}-
\frac{1}{\varepsilon^2}\delta(f-\varepsilon)
+\left[\log(2\pi\varepsilon)+\gamma-1\right]
\delta^{(2)}(f-\varepsilon)\biggr\}+
O(\varepsilon^2).
\label{9}
\end{equation}
\vspace{0.3 cm}

\item \underline{Random walk in frequency}:
\vspace{0.3 cm}
\begin{equation}
S(f)=\frac{h_4}{16\pi^4}\biggl\{
\frac{1}{f^4}-
\frac{2}{3\varepsilon^3}\delta(f-\varepsilon)
-\frac{2}{3\varepsilon}
\delta^{(2)}(f-\varepsilon)\biggr\}+
O(\varepsilon).
\label{10}
\end{equation}
\vspace{0.3 cm}

\item \underline{Flicker noise in frequency derivative}:
\vspace{0.3 cm}
\begin{equation}
S(f)=\frac{h_5}{32\pi^5}\biggl\{
\frac{1}{f^5}-
\frac{1}{2\varepsilon^4}\delta(f-\varepsilon)
-\frac{1}{4\varepsilon^2}
\delta^{(2)}(f-\varepsilon)
+\frac{1}{12}\left[\gamma+\log(2\pi\varepsilon)-\frac{1}{3}\right]
\delta^{(4)}(f-\varepsilon)\biggr\}+
O(\varepsilon^2).
\label{11}
\end{equation}
\vspace{0.3 cm}

\item \underline{Random walk in frequency derivative}:
\vspace{0.3 cm}
\begin{equation}
S(f)=\frac{h_6}{64\pi^6}\biggl\{
\frac{1}{f^6}-
\frac{2}{5\varepsilon^5}\delta(f-\varepsilon)
-\frac{2}{15\varepsilon^3}
\delta^{(2)}(f-\varepsilon)
\biggr\}+
O(\varepsilon).
\label{rww}
\end{equation}
\end{enumerate}
\vspace{0.3 cm}
Let us note that the coefficient $B_4(\varepsilon)\equiv 0$ in 
the expression (\ref{rww}) for spectrum of random walk in 
frequency derivative.
  
The expressions given above indicate that there is a strong concentration of 
infinite energy of the noise at
the lower cut-off frequency. As we have stressed already it is a specific
feature of the chosen model of the spectrum which appears because we do not 
know a real behavior
of spectrum while frequency is approaching to zero. Another remark is that 
a direct integration of any of the foregoing spectra with respect
to frequency from $f=\varepsilon$ to infinity 
(which may be erroneosly treated as the energy being stored in TOA residuals) 
gives zero value which may look surprising. However, 
it is worth noting that the entire energy presents in TOA residuals can 
be calculated
only after multiplication of the spectrum by the filter function (see section 6
below for more detail).
Therefore, calculation of the total energy of residuals is more complicated and
always gives a positive numerical value as it should be. Similar arguments 
can be used in
calculating variances of fitting parameters. For example, calculation of
variances of the first several spin-down parameters in frequency
domain may give a negative numerical value of the variance (Kopeikin 1999)
which is physically meaningful. The paradox is solved if we remember about
contribution of the non-stationary part of noise which always makes variances
of the parameters numerically positive (for more detail see (Kopeikin 1999)).
 
Pulsar timing observations can be used for estimation of strength and spectrum
of low frequency noise presents in TOA residuals. For this reason, development
of practically useful estimators of spectrum of noise are
required. We are not going to consider in the present paper the question 
about how to construct the best possible estimators. This subject has been 
enlightened by a number of other
researches (see, for instance, the papers of Deeter \& Boynton
(1982), Deeter(1984), Taylor 1991, Matsakis {\it al.} 1997). Our purpose is to 
study spectral dependence of TOA residuals and variances of 
fitting parameters which are used in real practice. In order to make clear 
what we are doing let us describe, first of all, the timing model we are 
dealing with.  

\section{Timing Model}

We consider a simplified, but still realistic model of arrival time
measurements of pulses from a pulsar in a binary system. It
is assumed that the orbit is circular, and
the pulsar rotates around its own axis with angular frequency $\nu _{p}$
which slows down due to the electromagnetic (or whatever) energy losses. It
is also taken into account that the orbital frequency of the binary system, $%
n_{b},$ and its projected semimajor axis, $x,$ have a
secular drift caused by radial acceleration of the
binary (Damour \& Taylor 1991, Bell \& Bailes 1996), its proper motion of  
in the sky (Kopeikin 1996), and emission of gravitational waves from the binary (Peters
\& Mathews 1963, Peters 1964) bringing about the gravitational radiation 
reaction force (Damour 1983a, Grishchuk \& Kopeikin 1983).

The moment ${\cal T}$ of emission of the ${\cal N}$-th pulsar's pulse
relates to the moment $t$ of its arrival measured at the infinite
electromagnetic frequency by the equations (Damour \& Taylor 1992, Kopeikin
1994, 1999): 
\begin{equation}
D\left[{\cal T}+x\sin \left( n_{b}{\cal T}+\sigma \right)\right]
 =t+\varphi_0(t)+\varphi_1(t),
\label{144}
\end{equation}
\begin{equation}
t=\tau^{\ast} +\Delta _{C}+\Delta _{R\odot }+\Delta _{\pi \odot }+\Delta _{E\odot
}+\Delta _{S\odot }.  \label{144a}
\end{equation}
We use the following notations:

\begin{itemize}
\item  ${\cal T}$ - pulsar time scale,

\item  $t$ - barycentric time at the barycenter of the Solar system,

\item  $\tau^{\ast} $ - topocentric time of observer,

\item  $\Delta _{C},$ $\Delta _{R\odot },$ $\Delta _{\pi \odot },$ $\Delta
_{E\odot },$ $\Delta _{S\odot }$ - clock and astrometric corrections (Taylor
\& Weisberg 1989, Doroshenko \& Kopeikin 1990, 1995) which one assumes to be
known precisely,

\item  $D$ - Doppler factor gradually changing due to the acceleration and
proper motion of the binary system in the sky \footnote
{$D=\frac{1+\frac{V_R}{c}}{\sqrt{\left(1-\frac{V^2}{c^2}\right)}}$, 
where $V_R$ and
$V$ are correspondingly the
relative radial and total velocities of the binary system barycentre with 
respect to the barycentre of the Solar system} 

\item  $\sigma $ - initial (constant) orbital phase,

\item  $n_{b}$ - orbital frequency $(n_{b}=2\pi /P_{b})$,

\item  $i$ - angle of inclination of the orbit to the line of sight,

\item  $x$ - projected semimajor axis $a_{p}$ of the pulsar's
orbit $(x=a_{p}\sin i/c)$,

\item  $c$ - speed of light,

\item  $\varphi_0 (t)$ - the gaussian noise of TOA measuring errors,

\item  $\varphi_1 (t)$ - low-frequency gaussian noise caused by the long-term 
instabilities of terrestrial clocks, effects in propagation of radio signals in
the interstellar medium and stochastic background of primordial gravitational
waves, etc.
\end{itemize}

The rotational phase of the pulsar is given by the polynomial in time 
\begin{equation}
{\cal N}(t)=\nu _{p}{\cal T}+\frac{1}{2}\stackrel{.}{\nu }_{p}{\cal T}^{2}+%
\frac{1}{6}\stackrel{..}{\nu }_{p}{\cal T}^{3}+\frac{1}{24}\stackrel{...}{%
\nu }_{p}{\cal T}^{4}+\frac{1}{120}\stackrel{....}{\nu }_{p}{\cal T}%
^{5}+\nu_p\varphi_2({\cal T})+O\left({\cal T}^6\right),   \label{155}
\end{equation}
where $\nu _{p},$ $\stackrel{.}{\nu }_{p},\stackrel{..}{\nu }_{p},$ {\it etc.%
} are pulsar's rotational frequency and its time derivatives all referred to
the epoch ${\cal T}=0,$ the term $O\left({\cal T}^6\right)$ denotes high order 
derivatives
of the rotational phase, and $\varphi_2({\cal T})$ is the intrinsic pulsar timing 
noise in either
rotational phase, frequency, or frequency derivative. Solving iteratively 
equation (\ref{144}) with respect to $%
{\cal T}$ and substituting ${\cal T}$ for the right hand side of equation (%
\ref{155}) gives a relationship between two observable quantities ${\cal N}$
and $t$: \vspace{0.3 cm} 
\begin{eqnarray}
{\cal N}(t)&=&{\cal N}_{0}+\nu t+{\frac{1}{2}}\stackrel{.}{\nu } t^2+
{\frac{1%
}{6}}\stackrel{..}{\nu }t^3+{\frac{1}{24}}\stackrel{...}{\nu }t^4\nonumber
+\\ &&\mbox{} {\frac{%
1}{120}}\stackrel{....}{\nu }t^5-  
\nu (x+\stackrel{.}{x}t+{\frac{1}{2}}\stackrel{..}{x}t^2+{\frac{1}{6}}%
\stackrel{...}{x}t^3)\sin (\sigma +n_{b}t+{\frac{1}{2}}\dot{n}_{b}t^2+ 
{%
\frac{1}{6}}\ddot{n}_{b}t^3)+\nu \epsilon (t),   
\label{177}
\end{eqnarray}\vspace{0.3 cm}
\begin{equation}
\epsilon(t) \stackrel{def}{=}\varphi_0(t)+\varphi_1(t)+\varphi_2(t),
\label{177aa}\vspace{0.3 cm}
\end{equation}
where ${\cal N}_{0}$ is the initial rotational phase of the pulsar $({\cal N}%
_{0}\simeq -\nu t_{0})$; $\nu ,$ $\stackrel{.}{\nu },$ $\stackrel{..}{\nu }%
,...$ are the pulsar's rotational frequency and its time derivatives at the
initial epoch $t_{0};$ $x,$ $\stackrel{.}{x},$ $\stackrel{..}{x},...$ are
the projected semimajor axis of the orbit and its time derivatives at the
epoch $t_{0};$ $\sigma ,$ $n_{b},$ $\dot{n}_{b},$ $\ddot{n}_{b},...$ are the
pulsar's orbital initial phase, orbital frequency and its time derivatives
at the epoch $t_{0}.$ Timing model (\ref{177}) accounts only for linear terms
which is enough for implication of least square method of fitting parameters to
the data. All non-linear residual terms of order $x^{2},
x\epsilon ,$ $%
\epsilon ^{2},$ $\dot{\nu}x$, $\ddot{\nu}x$, etc. are negligible and, for this
reason, 
have been omitted from (\ref{177}).

We assume that all observations of the binary pulsar are of a similar quality
and weight.
Then one defines the timing residuals $r(t)$ as a difference between the
observed number of the pulse, ${\cal N}^{obs},$ and the number ${\cal N}%
(t,\theta ),$ predicted on the ground of our best guess to the prior unknown
parameters of timing model (\ref{177}), divided by the pulsar's rotational
frequency $\nu $, that is\vspace{0.3 cm} 
\begin{equation}
r(t,\theta )=\frac{{\cal N}^{obs}-{\cal N}(t,\theta )}{\nu },\vspace{0.3 cm}
\label{188a}
\end{equation}\vspace{0.3 cm}
where $\theta =\{\theta _{a},a=1,2,...k\}$ denotes a set of $k$ measured
parameters $\left(k=14 \mbox{in the model} (\ref{177})\right)$ which are shown in 
Table
{\ref{tab:par}}. It is worth noting that hereafter we use for the reason of 
convinuence the time argument $u=n_b t$, that is the current orbital phase.
\begin{table*}
\begin{minipage}{140mm}
\centering
\scriptsize\caption{List of the basic functions and parameters used in the fitting 
procedure. Spin parameters $\delta {\cal N}_{0}, \delta {\nu },
\delta\stackrel{.}{\nu }, \delta\stackrel{..}{\nu }, 
\delta\stackrel{...}{\nu }, \delta\stackrel{....}{\nu },$ fit rotational 
motion of the pulsar around its own axis. Keplerian parameters $\delta{x},
\delta\sigma, \delta{n}_{b}$ fit the Keplerian orbital motion of the pulsar 
about barycentre of the binary system. Post-Keplerian parameters 
$\delta\stackrel{.}{x}, \delta\stackrel{..}{x}, \delta\stackrel{...}{x}, 
\delta\stackrel{.}{n}_{b}, \delta\stackrel{..}{n}_{b}$ fit small observable
deviations of the pulsar's orbit from the Keplerian motion caused by the
effects of General Relativity, radial acceleartion, and proper motion
of barycentre of the binary system with respect to the observer}
\label{tab:par}
\vspace{5 mm}
\begin{tabular}{|ll@{\hspace{5 cm}}ll|}
\hline \\ \\
Parameter &&& Fitting Function \\ \hline
&  \\ 
$\beta _{1}={{\frac{\delta {\cal N}_{0}}{\nu }}}$&&& $\psi _{1}(t)=1$ \\ 
&  \\ 
$\beta _{2}={\frac{1}{n_{b}}}{{\frac{\delta \nu }{\nu }}}$ &&& $\psi
_{2}(t)=u$ \\ 
&  \\ 
$\beta _{3}={\frac{1}{2n_{b}^{2}}}{{\frac{\delta \stackrel{.}{\nu }}{\nu }}}%
 $ &&& $\psi _{3}(t)=u^{2}$ \\ 
&  \\ 
$\beta _{4}={\frac{1}{6n_{b}^{3}}}{{\frac{\delta \stackrel{..}{\nu }}{\nu }}}%
$ &&& $\psi _{4}(t)=u^{3}$ \\ 
&  \\ 
$\beta _{5}={\frac{1}{24n_{b}^{4}}}{{\frac{\delta \stackrel{...}{\nu }}{\nu }%
}}$ &&& $\psi _{5}(t)=u^{4}$ \\ 
&  \\ 
$\beta _{6}={\frac{1}{120n_{b}^{5}}}{{\frac{\delta \stackrel{....}{\nu }}{%
\nu }}}$ &&& $\psi _{6}(t)=u^{5}$ \\ 
&  \\ 
$\beta _{7}=-\delta x\sin \sigma -\delta \sigma x\cos \sigma $ &&& $\psi
_{7}(t)=\cos u$ \\ 
&  \\ 
$\beta _{8}=-\delta x\cos \sigma +\delta \sigma x\sin \sigma $ &&& $\psi
_{8}(t)=\sin u$ \\ 
&  \\ 
$\beta _{9}={\frac{1}{n_{b}}}\left( -\delta \stackrel{.}{x}\cos \sigma
+\delta n_{b}x\sin \sigma \right) $ &&& $\psi _{9}(t)=u\sin u$ \\ 
&  \\ 
$\beta _{10}={\frac{1}{n_{b}}}\left( -\delta \stackrel{.}{x}\sin \sigma
-\delta n_{b}x\cos \sigma \right) $ &&& $\psi _{10}(t)=u\cos u,$ \\ 
&  \\ 
$\beta _{11}={\frac{1}{2n_{b}^{2}}}\left( -\delta \stackrel{..}{x}\sin
\sigma -\delta \stackrel{.}{n}_{b}x\cos \sigma \right) $ &&& $\psi
_{11}(t)=u^{2}\cos u$ \\ 
&  \\ 
$\beta _{12}={\frac{1}{2n_{b}^{2}}}\left( -\delta \stackrel{..}{x}\cos
\sigma +\delta \stackrel{.}{n}_{b}x\sin \sigma \right) $ &&& $\psi
_{12}(t)=u^{2}\sin u$ \\ 
&  \\ 
$\beta _{13}={\frac{1}{6n_{b}^{3}}}\left( -\delta \stackrel{...}{x}\cos
\sigma +\delta \stackrel{..}{n}_{b}x\sin \sigma \right) $ &&& $\psi
_{13}(t)=u^{3}\sin u$ \\ 
&  \\ 
$\beta _{14}={\frac{1}{6n_{b}^{3}}}\left( -\delta \stackrel{...}{x}\sin
\sigma -\delta \stackrel{..}{n}_{b}x\cos \sigma \right) $ &&& $\psi
_{14}(t)=u^{3}\cos u$ \\ \\
\hline  
\end{tabular}
\end{minipage}
\end{table*}
If a numerical value of the parameter $\theta _{a}$ coincides with its
true physical value $\hat{\theta}_{a}$, then the set of residuals would
represent a physically meaningful noise $\epsilon (t)$, {\it i.e. }
\begin{equation}
r(t,\hat{\theta})=\epsilon (t).
\end{equation}
In practice, however, the true values of parameters
are not attainable and we deal actually with their least square estimates $%
\theta _{a}^{*}.$ Therefore, observed residuals are fitted to the expression
which is a linear function of corrections to the estimates $\theta _{a}^{*}$
of a priori unknown true values of parameters $\hat{\theta}_{a}$. From a
Taylor expansion of the timing model in equation (\ref{177}), and the
fact that $r(t,\hat{\theta})=\epsilon (t)$ one obtains\vspace{0.3 cm} 
\begin{equation}
r(t,\theta ^{*})=\epsilon (t)-{\displaystyle \sum_{a=1}^{14}}\beta _{a}\psi _{a}(t,\theta
^{*})+O(\beta _{a}^{2}),  \label{199}
\end{equation}\vspace{0.3 cm}
where the quantities $\beta _{a}\equiv \delta \theta _{a}=\theta _{a}^{*}-%
\hat{\theta}_{a}$ are the corrections to the presently unknown true values of
parameters, and $\psi _{a}(t,\theta ^{*})=\left[ \frac{\partial {\cal N}}{%
\partial \theta _{a}}\right] _{\theta =\theta ^{*}}$ are basic fitting
functions of the timing model.

In the following it is more convenient to regard the increments $\beta _{a}$
as new parameters whose values are to be determined from the fitting
procedure. The parameters $\beta _{a}$ and fitting functions are summarized in
Table {\ref{tab:par}} with asterisks omitted and time $t$ is replaced for convenience by
the function $u=n_{b}t$ which is the current value of orbital phase.
We restrict the model to $14$
parameters since in practice only the first several parameters of the model
are significant in fitting to the rotational and orbital phases over the
available time span of observations. 

Let us introduce auxilary functions $\Xi_a(t)$ defined according to the
formula:
\vspace{0.3 cm}
\begin{equation}
\label{ert}
\Xi_a(t)=\displaystyle{\sum_{b=1}^{14}}L_{ab}^{-1}\psi_b(t),
\end{equation}
where the matrix of information 
\vspace{0.3 cm} 
\begin{equation}
L_{ab}({\it T})={\displaystyle \sum_{i=1}^{mN}}\psi _{a}(t_{i})\psi _{b}(t_{i}),\vspace{0.3 cm}
\label{111a}
\end{equation}\vspace{0.3 cm}
the matrix $L_{ab}^{-1}$ is its inverse, and ${\it T}=NP_{b}$ is a total
span of observational time.
Functions $\Xi(t)$ are called (Deeter 1984) the dual ones to $\psi_a(t)$ 
because of the cross-orthonormality condition is hold:
\vspace{0.3 cm}
\begin{equation}
\label{oik}
\displaystyle{\sum_{i=1}^{mN}}\Xi_a(t_i)\psi_b(t_i)=\delta_{ab}.
\end{equation}
Now suppose that we measure $m$ equally spaced and comparably accurate
arrival times each orbit for a total of $N$ orbital revolutions, so we have $%
mN$ residuals $r_{i}\equiv r(t_{i}),$ $i=1,...,mN.$ Standard least squares
procedure (Bard 1974) gives the best fitting solution for estimates
of the parameters $\beta _{a}$ \vspace{0.3 cm} 
\begin{equation}
\beta _{a}({\it T})={\displaystyle \sum_{i=1}^{mN}}\Xi_a(t_i)
\epsilon (t_{i}),\qquad a=1,...,14.\vspace{0.3 cm}  \label{111}
\end{equation}
\vspace{0.3 cm}  

Let the angular brackets denote an ensemble average over many
realizations of the observational procedure. Hereafter, we assume that 
the ensemble average of the noise $\epsilon(t)$ is equal to
zero. 
Hence, the mean value of any parameter $\beta_{a}$ is equal to zero as
well, {\it i.e.}
\begin{equation}
<\epsilon(t)> =0 \hspace{1 cm}\longrightarrow \hspace{1 cm} <\beta _{a}>=0.
\label{mean}
\end{equation}
The covariance matrix 
$M_{ab}\equiv $ $<\beta _{a}\beta _{b}>$ of the
parameter estimates is now given by the expression \vspace{0.3 cm} 
\begin{equation}
M_{ab}({\it T})=
{\displaystyle \sum_{i=1}^{mN}}{\displaystyle \sum_{j=1}^{mN}}
\Xi _{a}(t_{i})\Xi
_{b}(t_{j})R(t_{i},t_{j}) ,\vspace{0.3 cm}  
\label{112}
\end{equation}\vspace{0.3 cm}
where $R(t_{i},t_{j})=$ $<\epsilon (t_{i})\epsilon (t_{j})>$ is the
autocovariance function of the stochastic process $\epsilon (t)$.
The covariance matrix is
symmetric (that is, $M_{ab}=M_{ba}$), 
elements of its main diagonal give
variations (or dispersions) of measured parameters 
$\sigma_{\beta _{a}}\equiv M_{aa}$ = $<\beta_{a}^{2}>$, and the off-diagonal 
terms represent the degree of statistic covariance (or correlation) between 
them. Covariance matrix consists of two additive components $M_{ab}^{+}$ and
$M_{ab}^{-}$ describing correspondingly contributions from non-stationary and
stationary parts of autocovariance function $R(t_i,t_j)$. Explicit expressions
for the matrix $M_{ab}$ can be found in the paper (Kopeikin 1988) wherein we
have done all calculations in the time domain. Only $M_{ab}^{-}$ admits
transformation to the frequency domain which will
be discussed in subsequent sections.

Subtraction of the adopted model from the observational data leads to the
residuals which are dominated by the random fluctuations only. An expression
for the mean-square residuals after subtracting the best-fitting solution
for the estimates (\ref{111}) is given by the formula \vspace{0.3 cm} 
\begin{equation}
<r^{2}({\it T})>=\frac{1}{mN}{\displaystyle \sum_{i=1}^{mN}}%
{\displaystyle \sum_{j=1}^{mN}}K(t_{i},t_{j})R(t_{i},t_{j}),\vspace{0.3 cm}  
\label{112a}
\end{equation}\vspace{0.3 cm}
where the function 
\begin{equation}\vspace{0.3 cm}
K(t_{i},t_{j})=\delta _{ij}-
{\displaystyle \sum_{a=1}^{14}}\Xi_{a}(t_{i})\psi _{a}(t_{j}),\vspace{0.3 cm}  \label{112b}
\end{equation}\vspace{0.3 cm}
is called the filter function (Blandford {\it et al.} 1984). 
We have proved (Kopeikin 1999) that the post-fit residuals depend only on the 
stationary part of the noise 
\begin{equation}\vspace{0.3 cm}
<r^{2}({\it T})>=\frac{1}{mN}%
{\displaystyle \sum_{i=1}^{mN}}%
{\displaystyle \sum_{j=1}^{mN}}K(t_{i},t_{j})R^{-}(t_{i},t_{j})=
-\frac{1}{mN}{\displaystyle \sum_{a=1}^{14}}{\displaystyle 
\sum_{b=1}^{14}}L_{ab}^{-1}\left[
{\displaystyle \sum_{i=1}^{mN}}{\displaystyle \sum_{j=1}^{mN}}
\psi _{a}(t_{i})\psi
_{b}(t_{j})R^{-}(t_{i},t_{j})\right] 
. \label{vlbi}
\end{equation}\vspace{0.3 cm}
For this reason, methods of spectral analysis in frequency domain can be 
applied for analyzing residuals without any restriction. Let us note that the
explicit dependence of TOA residuals on the total span of observations contains
in (Kopeikin 1988).
      
\section{Fourier Transform of Fitting Functions}

We define the Fourier transform of the fitting functions $\psi_a(t)$ as
\vspace{0.3 cm}
\begin{equation}
\label{summa}
\tilde{\Psi}_a(f,m,N)=\displaystyle{\sum_{j=1}^{mN}}\psi_a(t_j)
\exp(-2\pi i f t_j),
\end{equation}
\vspace{0.3 cm}
where $f$ is the Fourier frequency measured in units being inversly
proportional to units of measurement of time $t$. We measure time in units 
of orbital phase $u=n_b t$, that is in
radians. Then the frequency $\omega =2\pi f$ is dimensionless and 
measured in units of orbital frequency $n_b$. One notes the Fourier transfrom
of the fitting functions depends on the Fourier frequency $f$, total amount of 
orbital revolutions $N$, and frequency of observations $m$.   

When the total amount of observational points, $mN$, is large we can 
approximate the sum (\ref{summa}) by the integral (Kopeikin 1999)
\vspace{0.3 cm}
\begin{equation}
\label{integ}
\tilde{\Psi}_a(\omega,m,N)=\frac{m}{2\pi}\tilde{\psi}_a(\omega,N)
\end{equation}
\vspace{0.3 cm}
\begin{equation}
\label{in}
\tilde{\psi}_a(\omega,N)=
\displaystyle{\int_{-\pi N}^{\pi N}}\psi_a(u) \exp(-i \omega u) du.
\label{ft1}
\end{equation}
We note that $\tilde{\psi}_a(-\omega)=
\tilde{\psi}_a^{\ast}(\omega)$, where the asterisk denotes a complex 
conjugation.
Replacing the sum over observational points by the integral with respect to
time (or orbital phase) is equivalent to the case of continuous observations.

The following formulae are also of use in practical computations:
\vspace{0.3 cm}
\begin{equation}
\hspace{0.3 cm}\tilde{\psi}_a(\omega,N)=\left\{\begin{array}{ll}
2\displaystyle{\int_{0}^{\pi N}}\psi_a(u)
\cos(\omega u) du,\hspace{0.3 cm} \mbox{if index $a=1,3,5,...$}\nonumber\\ \\ \nonumber
-2i\displaystyle{\int_{0}^{\pi N}}\psi_a(u)
\sin(\omega u) du,\hspace{0.3 cm}\mbox{if index $a=2,4,6,...$.}
\end{array}\right.
\label{ft3}
\end{equation}
\vspace{0.3 cm}
These expressions shows that the fitting functions with odd indices are real
and those with even ones do complex. 

Let us introduce notations - ${\rm T}=\pi N$, $z=\omega T$.
The Fourier transform of fitting functions takes the form:
\begin{eqnarray}
\tilde{\psi}_1(\omega)&=&2{\rm T}\tilde{\phi}_1(z),
\label{f1}
\end{eqnarray}
\begin{eqnarray}
\tilde{\psi}_2(\omega)&=&2i{\rm T}^2\tilde{\phi}_2(z),
\label{f2}
\end{eqnarray}
\begin{eqnarray}
\tilde{\psi}_3(\omega)&=&2{\rm T}^3\tilde{\phi}_3(z),
\label{f3}
\end{eqnarray}
\begin{eqnarray}
\tilde{\psi}_4(\omega)&=&2i{\rm T}^4\tilde{\phi}_4(z),
\label{f4}
\end{eqnarray}
\begin{eqnarray}
\tilde{\psi}_5(\omega)&=&2{\rm T}^5\tilde{\phi}_5(z),
\label{f5}
\end{eqnarray}
\begin{eqnarray}
\tilde{\psi}_6(\omega)&=&2i{\rm T}^6\tilde{\phi}_6(z),
\label{f6}
\end{eqnarray}
\begin{eqnarray}
\tilde{\psi}_7(\omega)&=&{\rm T}\left[\tilde{\phi}_1(z+{\rm T})+
\tilde{\phi}_1(z-{\rm T})\right],
\label{f7}
\end{eqnarray}
\begin{eqnarray}
\tilde{\psi}_8(\omega)&=&i{\rm T}\left[\tilde{\phi}_1(z+{\rm T})-
\tilde{\phi}_1(z-{\rm T})\right],
\label{f8}
\end{eqnarray}
\begin{eqnarray}
\tilde{\psi}_9(\omega)&=&{\rm T}^2\left[\tilde{\phi}_2(z-{\rm T})
-\tilde{\phi}_2(z+{\rm T})\right],
\label{f9}
\end{eqnarray}
\begin{eqnarray}
\tilde{\psi}_{10}(\omega)&=&i{\rm T}^2\left[\tilde{\phi}_2(z-{\rm T})
+\tilde{\phi}_2(z+{\rm T})\right],
\label{f10}
\end{eqnarray}
\begin{eqnarray}
\tilde{\psi}_{11}(\omega)&=&{\rm T}^3\left[\tilde{\phi}_3(z+{\rm T})+
\tilde{\phi}_3(z-{\rm T})\right],
\label{f11}
\end{eqnarray}
\begin{eqnarray}
\tilde{\psi}_{12}(\omega)&=&i{\rm T}^3\left[\tilde{\phi}_3(z+{\rm T})-
\tilde{\phi}_3(z-{\rm T})\right],
\label{f12}
\end{eqnarray}
\begin{eqnarray}
\tilde{\psi}_{13}(\omega)&=&{\rm T}^4\left[\tilde{\phi}_4(z-{\rm T})-
\tilde{\phi}_4(z+{\rm T})\right],
\label{f13}
\end{eqnarray}
\begin{eqnarray}
\tilde{\psi}_{14}(\omega)&=&i{\rm T}^4\left[\tilde{\phi}_4(z-{\rm T})+
\tilde{\phi}_4(z+{\rm T})\right].
\label{f14}
\end{eqnarray}
\vspace{0.5 cm}
Functions $\phi_a(z)$ ($a=1,2,...,6)$ have the following form: 
\begin{eqnarray}
\tilde{\phi}_1(z)&=&\frac{\sin z}{z},
\label{ff1}
\end{eqnarray}

\begin{eqnarray}
\tilde{\phi}_2(z)&=&
\frac{\cos z }{z}-
\frac{\sin z}{z^2},
\label{ff2}
\end{eqnarray}

\begin{eqnarray}
\tilde{\phi}_3(z)&=&\frac{\sin z}{z}+
\frac{2 \cos z }{z^2}-
\frac{2 \sin z }{z^3},
\label{ff3}
\end{eqnarray}

\begin{eqnarray}
\tilde{\phi}_4(z)&=&\frac{\cos z }{z}-
\frac{3 \sin z }{z^2}-
\frac{6\cos z}{z^3}+
\frac{6 \sin z}{z^4},
\label{ff4}
\end{eqnarray}

\begin{eqnarray}
\tilde{\phi}_5(z)&=&\frac{\sin z}{z}+
\frac{4\cos z}{z^2}-
\frac{12\sin z}{z^3}-
\frac{24\cos z }{z^4}+
\frac{24\sin z }{z^5},
\label{ff5}
\end{eqnarray}

\begin{eqnarray}
\tilde{\phi}_6(z)&=&\frac{\cos z}{z}-
\frac{5\sin z }{z^2}-
\frac{20\cos z }{z^3}+
\frac{60\sin z}{z^4}+
\frac{120\cos z }{z^5}-
\frac{120\sin z}{z^6}.
\label{ff6}
\end{eqnarray}

\begin{figure*}
\centerline{\psfig{figure=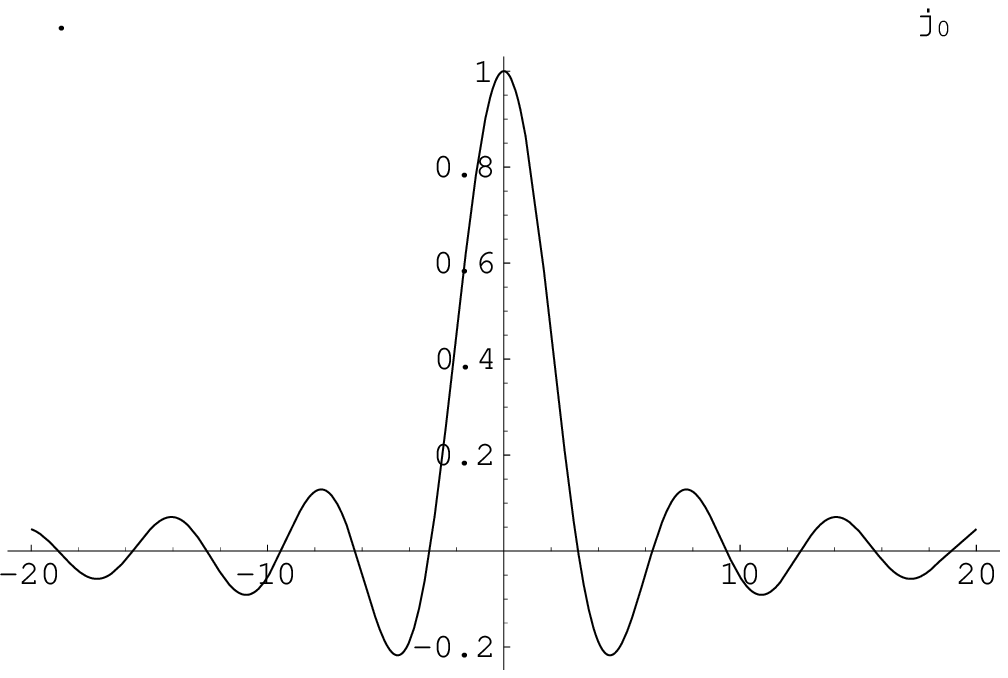,angle=0,height=5cm,width=8cm}
\hspace{1 cm}
\psfig{figure=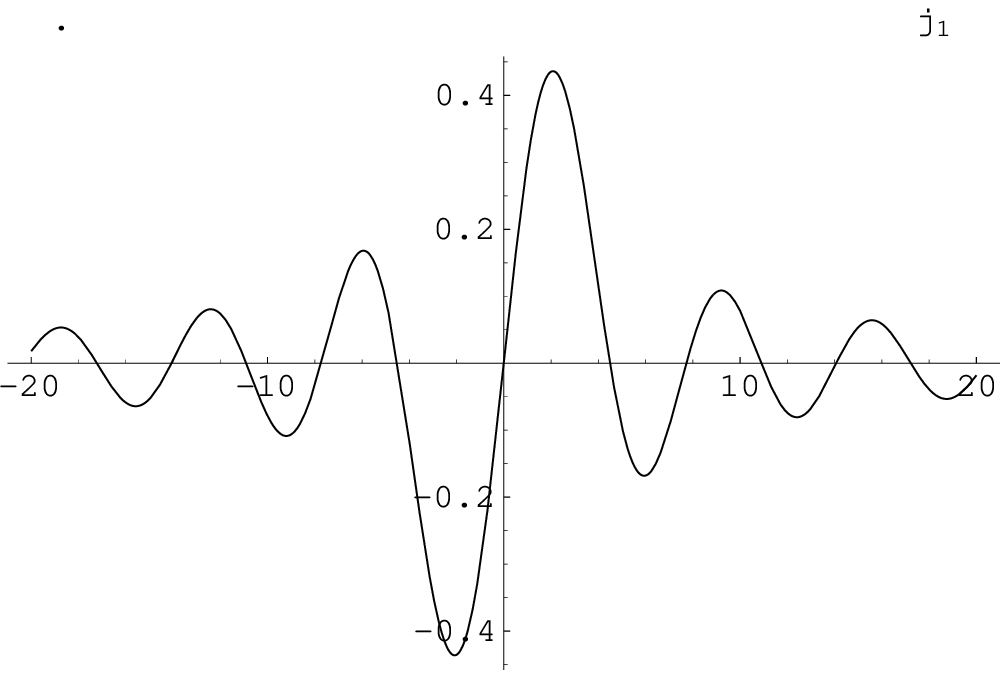,angle=0,height=5cm,width=8cm}}\vspace{1.5 cm}
\centerline{\psfig{figure=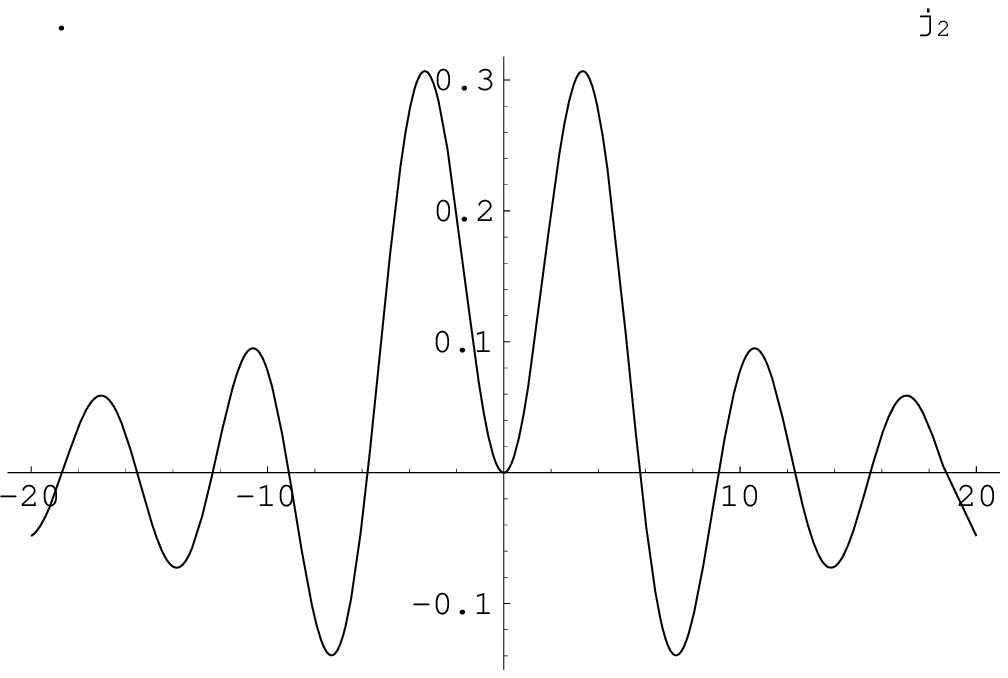,angle=0,height=5cm,width=8cm}
\hspace{1 cm}
\psfig{figure=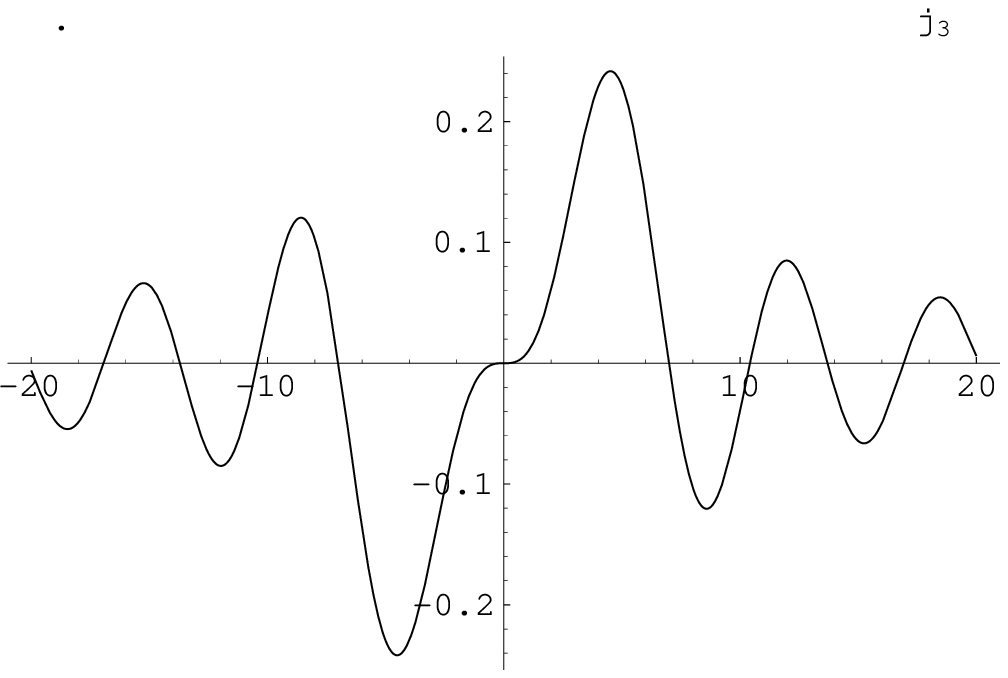,angle=0,height=5cm,width=8cm}}\vspace{1.5 cm}
\centerline{\psfig{figure=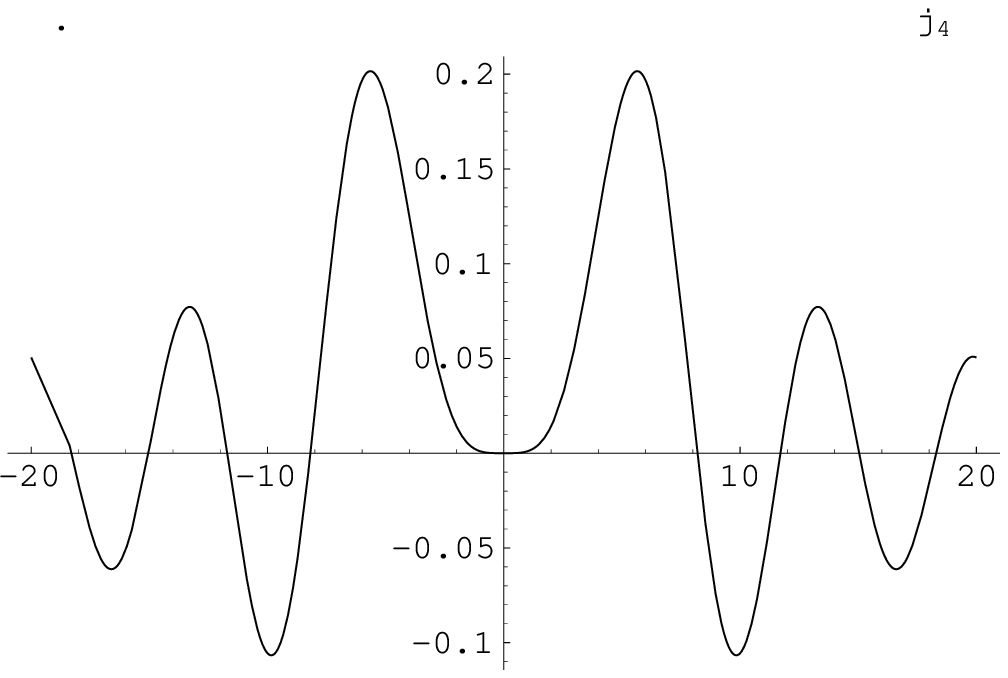,angle=0,height=5cm,width=8cm}
\hspace{1 cm}
\psfig{figure=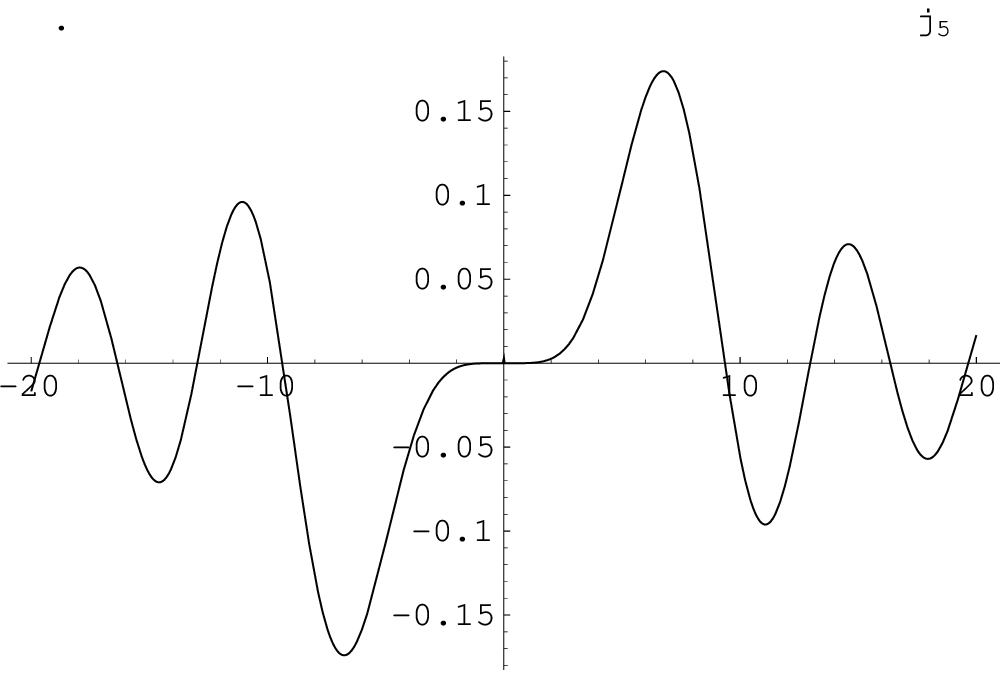,angle=0,height=5cm,width=8cm}}\vspace{1.5 cm}
\caption{Plots of the Fourier transforms of the sperical Bessel functions $j_0(z)$,
$j_1(z)$,\dots, $j_5(z)$ in terms of the variable $z=\omega{\rm T}$.
Ciclic frequency $\omega$ is measured in units of orbital frequency $n_b$. 
Amplitude of the transform is normalized to unity.}
\label{phi1}
\end{figure*}

Actually, it is more convenient to use in what follows the spherical Bessel
functions $j_a(z)$ defined as (Korn \& Korn 1968, section 21.8-8)
\begin{equation}\label{bessel}
j_a(z)=z^a\left(-\frac{1}{z}\frac{d}{dz}\right)^a \frac{\sin
z}{z},\quad (a=0,1,2,...)
\end{equation}
Plots of the functions $j_0(z)$, $j_1(z)$,..., 
$j_5(z)$ are displayed in Figure (\ref{phi1}). The functions have a different
behavior near the point $z=0$\footnote{The function
$j_a(z)=\frac{z^a}{(2a+1)!!}+O(z^{a+2})$ for $z\ll 1$ ($a=0,1,2,...$).}, and 
then oscillate with monotonically
decreasing amplitude. Asymptotic expansion of the sperical Bessel functions for
large values of the variable $z$ are
given by the formula 
\begin{equation}
j_a(z)\approx \frac{\sin\left(z-\displaystyle{\frac{\pi a}{2}}\right)}{z}.
\end{equation} 
It is worth emphasizing that
the maximal value of any of these functions can not be large than 1.

Fitting functions $\tilde\phi_a(z)$ being 
expressed in terms of the spherical Bessel
functions assume the form
\begin{eqnarray}
\tilde\phi_1(z)&=&j_0(z),\\\nonumber\\
\tilde\phi_2(z)&=&-j_1(z),\\\nonumber\\
\tilde\phi_3(z)&=&\frac{1}{3}j_0(z)-\frac{2}{3}j_2(z),\\\nonumber\\
\tilde\phi_4(z)&=&-\frac{3}{5}j_1(z)+\frac{2}{5}j_3(z),\\\nonumber\\
\tilde\phi_5(z)&=&\frac{1}{5}j_0(z)-\frac{4}{7}j_2(z)+\frac{8}{35}j_4(z),\\\nonumber\\
\tilde\phi_6(z)&=&-\frac{3}{7}j_1(z)+\frac{4}{9}j_3(z)-\frac{8}{63}j_5(z).
\end{eqnarray}
\section{Fourier Transform of the Covariance Matrix} 

In order to calculate the covariance matrix we need to know the 
Fourier transform of the dual functions 
$\Xi_a(t)$. The transform is defined in accordance with definition (\ref{ert}) 
of the dual functions and takes the form 
\vspace{0.3 cm}
\begin{equation}
\tilde{\Xi}_a(f,m,N)=\displaystyle{
\sum_{c=1}^{14}}L_{ac}^{-1}\tilde{\Psi}_c(f,m,N),
\label{xi}
\end{equation}
and the cross-orthonormal condition in frequency domain is given by the
integral
\vspace{0.3 cm}
\begin{equation}
\displaystyle{\int_{0}^{\infty}}\tilde{\Xi}_a(f,m,N)\tilde{\Psi}_b(f,m,N)df=
\frac{1}{2}\delta_{ab}.
\label{dual}
\end{equation}

In the limit of continuous observations it is convinuent to 
introduce matrix $C_{ab}=\frac{2\pi}{m}L_{ab}$ instead of the matrix $L_{ab}$.
Explicit expression for the matrix $C_{ab}$ is given by the integral:
\vspace{0.3 cm}
\begin{equation}
C_{ab}=\displaystyle{\int_{-\pi N}^{\pi N}}\psi_a(u) \psi_b(u) du,
\end{equation}
and the result of calculation of the integral is given in the paper (Kopeikin
1999, Tables 5,6).
Then we have  $L_{ab}^{-1}=\frac{2\pi}{m}C_{ab}^{-1}$ and the dual function
$\tilde{\Xi}(f)$ can be recast as
\begin{equation}
\tilde{\Xi}_a(f,N)=\displaystyle{
\sum_{b=1}^{14}}C_{ab}^{-1}\tilde{\psi}_b(f,N).
\label{gfl}
\end{equation}
\vspace{0.3 cm}
Hence, comparing the eq. (\ref{gfl}) with (\ref{xi}) one concludes that 
in the limit of continuous observations the Fourier transfrom of the 
dual functions depend only on the
Fourier frequency and total amount of orbital revolutions as it was expected.
It is more insightful to express the dual functions (\ref{gfl}) in terms of 
the spherical Bessel functions (\ref{bessel}). The expressions obtained are
rather unwieldy and, for this reason they are given in Appendix A. Plots of the
Fourier transform of the spherical Bessel functions are given in Appendix D. 

Making use of definition of the Fourier transforms of stationary part of 
autocovariance function (\ref{2}) and the dual functions (\ref{xi})
we obtain the Fourier transform of stationary part of the covariance matrix
$M_{ab}^{-}(m,N)$
\begin{equation}
M_{ab}^{-}=\displaystyle{\int_{\varepsilon}^{\infty}}S(f)H_{ab}(f,m,N)df,
\label{mab}
\end{equation}
where $H_{ab}(f,m,N)$ is the transfer function given by the expression 
\begin{equation}
H_{ab}(f,m,N)=
\tilde{\Xi}_a(f,m,N)\tilde{\Xi}_b^{\ast}(f,m,N)+
\tilde{\Xi}_a^{\ast}(f,m,N)\tilde{\Xi}_b(f,m,N),
\label{h}
\end{equation}
and asterisk denotes a complex conjugation. For numerical computations of 
$M_{ab}^{-}$ the next formula can be used in practice
\vspace{0.3 cm}
\begin{eqnarray}
M_{ab}^{-}(m,N)&=&\frac{h_n}{(2\pi)^n}\int_{\varepsilon}^{\Lambda}H_{ab}
(f,m,N)f^{-n}df+\\\nonumber\\\nonumber&&
\frac{1}{2}h_n \left[B_0(\varepsilon)H_{ab}(\varepsilon,m,N)+
\varepsilon^2 B_2(\varepsilon) H_{ab}^{(2)}(\varepsilon,m,N)+
\varepsilon^4 B_4(\varepsilon) H_{ab}^{(4)}(\varepsilon,m,N)+...\right],
\end{eqnarray}
where derivatives of $H_{ab}$ are taken with respect to the Fourier frequency, 
ellipses denote terms of negligible influence on the result of the
computation, and $\Lambda$ is the upper cut-off frequency arising from the
sampling theorem and inversly proportional to the minimal time between
subsequent observational sessions. 

\section{Fourier Transform of Residuals}

Fourier transform of timing 
residuals is obtained from eqs. (\ref{2}),
(\ref{vlbi}), and (\ref{ft1}). This yields:\vspace{0.3 cm}
\begin{equation}
<r^2> =
2\displaystyle{\int_{\varepsilon}^{\infty}}S(f)K(f,m,N)df,
\label{resid}
\end{equation}
\vspace{0.3 cm}
where $K(f,m,N)$ is the Fourier transform of the filter function (\ref{112b})
\begin{equation}
K(f,m,N)=1-\frac{1}{2mN}\displaystyle{\sum_a^{14}}
\left[\tilde{\Xi}_a(f,m,N)\tilde{\Psi}_a^{\ast}(f,m,N)+
\tilde{\Xi}_a^{\ast}(f,m,N)\tilde{\Psi}_a(f,m,N)\right].
\label{filt}
\end{equation}
In the limit of continuous observations there is no dependence on the frequency
of observations, $m$, so that one obtains
\begin{equation}
K(f,N)=1-\frac{1}{4{\rm T}}\displaystyle{\sum_a^{14}}
\left[\tilde{\Xi}_a(f,N)\tilde{\psi}_a^{\ast}(f,N)+
\tilde{\Xi}_a^{\ast}(f,N)\tilde{\psi}_a(f,N)\right].
\label{filter}
\end{equation}

Plot of the Fourier transform (\ref{filter}) of the filter function 
$K(f)$ is shown in Appendix E for different amount of orbital revolutions 
$N$. It is approximately equal to 1 until
frequency is higher than $1/T$, and rapidly decreases its amplitude 
as frequency approaches to zero. We also note that the curve of the 
Fourier transform clearly shows the additional dip
near the orbital frequency. The dips near zero and orbital frequencies are 
getting narrower as a number of observational points increases. 

\section{Spectral Sensitivity of Timing Observations of 
Millisecond and Binary Pulsars}

Analytical expressions and graphical representations of Fourier transforms of
dual functions and timing residuals help us to understand in more detail
spectral sensitivity of single and binary pulsars to different frequency bands
in spectral decomposition of noise.
First of all, let consider behavior of Fourier transform of fitting functions
near zero and orbital frequencies.

It is easy to confirm after making use of Taylor expansion of exponential
function in (\ref{ft1}) near $\omega\simeq 0$ that\vspace{0.3 cm}
\begin{equation}
\tilde{\psi}_a(\omega)= \left\{\begin{array}{ll}
C_{a1}-\frac{1}{2}\omega^2 C_{a3}+\frac{1}{24}\omega^4 C_{a5}+\omega^6
p_{a},&\mbox{if a=1,3,5,...}\\ \\
i\left(
-\omega C_{a2}+\frac{1}{6}\omega^3 C_{a4}-\frac{1}{120}\omega^5 C_{a6}+\omega^7
p_{a}\right),& \mbox{if a=2,4,6,...,}
\end{array}\right.
\label{expanzero}
\end{equation}\vspace{0.3 cm}
where $p_a$ is a residual term depending only on the total amount of orbital
revolutions, $N$.
Taylor expansion of Fourier transform of fitting functions near the orbital
frequency yields:\vspace{0.3 cm}
\begin{equation}
\tilde{\psi}_a(\omega)= \left\{\begin{array}{ll}
C_{a7}+(\omega-1) C_{a9}-\frac{(\omega-1)^2}{2} C_{a.11}-
\frac{(\omega-1)^3}{6} C_{a.13}+(\omega-1)^4 
q_{a},&\mbox{if a=1,3,...}\\ \\
i\left[
-C_{a8}+(\omega-1) C_{a.10}+\frac{(\omega-1)^2}{2} C_{a.12}-
\frac{(\omega-1)^3}{6} C_{a.14}+(\omega-1)^4 q_{a}\right],& \mbox{if a=2,4,...},
\end{array}\right.
\label{expanorb}
\end{equation}
where $q_a$ is a residual term depending only on the total amount of orbital
revolutions, $N$ (let us remind that frequency is measured in units 
of orbital frequency $n_b$). Applying to Eqs.
(\ref{expanzero})-(\ref{expanorb}) definition of the dual functions
(\ref{gfl}) in the limit of continuos observations yields asymptotic behavior
of the dual functions
\vspace{0.3 cm}
\begin{equation}
\tilde{\Xi}_a(\omega)= \left\{\begin{array}{ll}
\delta_{a1}-\frac{1}{2}\omega^2 \delta_{a3}+\frac{1}{24}\omega^4 \delta_{a5}+
\omega^6 P_{a},&\mbox{if a=1,3,5,...}\\ \\
i\left(
-\omega \delta_{a2}+\frac{1}{6}\omega^3 \delta_{a4}-\frac{1}{120}\omega^5
\delta_{a6}+\omega^7 P_{a}\right)
,& \mbox{if a=2,4,6,...,}
\end{array}\right.
\label{dfzero}
\end{equation}\vspace{0.3 cm}
near zero frequency, and \vspace{0.3 cm}
\begin{equation}
\tilde{\Xi}_a(\omega)= \left\{\begin{array}{ll}
\delta_{a7}+(\omega-1) \delta_{a9}-\frac{(\omega-1)^2}{2} C_{a.11}-
\frac{(\omega-1)^3}{6} 
\delta_{a.13}+(\omega-1)^4 Q_{a},&\mbox{if a=1,3,...}\\ \\
i\left[-\delta_{a8}+(\omega-1) \delta_{a.10}+\frac{(\omega-1)^2}{2} 
\delta_{a.12}-
\frac{(\omega-1)^3}{6} \delta_{a.14}+(\omega-1)^4 Q_{a}\right],& 
\mbox{if a=2,4,...},
\end{array}\right.
\label{dforb}
\end{equation}
near the orbital frequency, where $P_a$ and $Q_a$ are residual terms. 
Table 2 shows the asymtotic behavior of the residual terms of the dual 
functions.

Now we can study asymptotic behaviour of filter function $K(f)$ defined by eq.
(\ref{filter}). Taking into account the fact that
$\sum_{b=1}^{14} C_{ab}^{-1}C_{bc}=\delta_{ac}$ - the unit matrix, we
obtain\vspace{0.3 cm}
\begin{equation}
K(f) = \left\{\begin{array}{ll}
{\rm B_0} \cdot \omega^6, &\mbox{when $\omega \rightarrow 0$}\\ \\
{\rm B_1} \cdot (\omega-1)^4, &\mbox{when $\omega \rightarrow 1$}
\end{array}\right.
\label{tfil}
\end{equation}\vspace{0.3 cm}
where $B_0$ and $B_1$ are numerical constants.
Such specific behavior of the filter function $K(f)$ significantly reduces the amount of
detected red noise below the cut-off frequency $f_c \simeq \alpha_c {\rm
T}^{-1}$ and in the frequency band $1-\alpha_b {\rm T}^{-1} \leq f 
\leq 1+\alpha_b {\rm T}^{-1}$ lying near the orbital frequency. Here constant
coefficents $\alpha_c$ and $\alpha_b$ can be determined by means of comparision
of calculations of mean value of timing residuals in time and frequency
domains (Kopeikin 1997a). Low frequencies
of noise power spectrum are fitted away by the polynomial fit for the spin-down
parameters of the observed pulsar. Frequencies being close to the orbital one
are fitted away by the fit for orbital parameters of the pulsar. Amount of
noise power remained in timing residuals 
after completion of fitting procedure is estimated by the expression\vspace{0.3
cm}
\begin{equation}
< r^2 > = 
2\displaystyle{\int_{\frac{\alpha_c}{\rm T}}^
{1-\frac{\alpha_b}{\rm T}}}S(f)df+
2\displaystyle{\int_{1+\frac{\alpha_b}{\rm T}}^{\infty}}S(f)df =
\frac{2\alpha_c^{1-n}{\rm T}^{n-1}}{(2 \pi)^n(n-1)}+
\frac{4\alpha_b}{(2 \pi)^n {\rm
T}}+O\left(\frac{1}{{\rm T}^3}\right).
\label{uuu}
\end{equation}\vspace{0.3 cm}
The second term in the right hand side of 
eq. (\ref{uuu}) shows amount of noise absorbed by fitting
orbital parameters. It is negligibly small comparatively with first term in the
right hand side and
can be not taken into account in practice. Hence, we declare that the post-fit
timing residuals can be used for estimation of amount of red noise and its
spectrum in frequency band just from $\alpha_c {\rm T}^{-1}$ up to infinity,
irrespectively of whether the pulsar is binary or not. This reasoning puts on
firm ground the estimates of spectral window of timing observations
and cosmological parameter $\Omega_g$, characterizing energy density of
stochastic gravitational waves in early universe, made by Kaspi {\it et al.}
(1994) and Camilo {\it et al.} (1994) using observations of binary pulsars PSR
B1855+09 and PSR J1713+0747 respectively. 

Analysis of spectral sensitivity of estimates of variances of spin-down and
orbital parameters is more cumbersome.
We are interested in which frequencies give the biggest contribution to the
variances. This is important to know, for example, in the event of using
variances of certain orbital parameters for setting the fundamental upper 
limit on $\Omega_g$ (Kopeikin 1997a, Kopeikin \& Wex 1999). 
Analitic calculations reveal the leading terms in asymptotic 
expansions of the dual functions near zero frequencies:     
\begin{table*}
\begin{minipage}{130mm}
\centering
\caption{Asymptotic behavior of residual terms of 
the dual functions $\tilde{\Xi}_a$ near zero
and orbital frequencies. Constant ${\rm h}=\cos{\rm T}=(-1)^N$.}
\label{tab:11}
\vspace{3 mm}
\begin{tabular}{|ccc|}
\hline \\ \\
Dual function &Residual term $P_a$&Residual term $Q_a$ \\ \hline
&  \\ 
$\tilde{\Xi}_{1}$&$-\frac{1}{33264} {\rm T}^6$&$
\frac{19}{56}
{\rm T}^2{\rm h}$ \\ 
&  \\ 
$\tilde{\Xi}_{2}$&$\frac{1}{61776} {\rm T}^6$&
$-\frac{1}{8}{\rm T}^2 {\rm h}$\\ 
&  \\ 
$\tilde{\Xi}_{3}$&$\frac{1}{1584}{\rm T}^4$&
$-\frac{17}{4}{\rm h}$\\ 
&  \\ 
$\tilde{\Xi}_{4}$&$-\frac{1}{6864} {\rm T}^4$&
$-\frac{3}{4} {\rm h}$\\ 
&  \\ 
$\tilde{\Xi}_{5}$&$-\frac{1}{528}{\rm T}^2$&
$-\frac{45}{8}\frac{{\rm h}}{{\rm T}^2}$\\ 
&  \\  
$\tilde{\Xi}_{6}$&$\frac{1}{3120}{\rm T}^2$&
$\frac{33}{40}\frac{{\rm h}}{{\rm T}^2} $\\ 
&  \\
$\tilde{\Xi}_{7}$ &$-\frac{1}{165}{\rm T}^4{\rm h}$&
$-\frac{1}{280} {\rm T}^4$ \\ 
&  \\ 
$\tilde{\Xi}_{8}$ &$\frac{1}{45045}{\rm T}^6{\rm h}$&
$\frac{1}{280} {\rm T}^4$ \\ 
&  \\ 
$\tilde{\Xi}_{9}$ &$-\frac{1}{693}{\rm T}^4{\rm h}$&
$-\frac{1}{28}{\rm T}^2$ \\ 
&  \\  
$\tilde{\Xi}_{10}$ &$-\frac{1}{273}{\rm T}^4{\rm h}$&
$\frac{1}{28} {\rm T}^2$ \\ 
&  \\ 
$\tilde{\Xi}_{11}$ &$-\frac{19}{693}{\rm T}^2{\rm h}$&
$\frac{1}{28} {\rm T}^2$ \\ 
&  \\
$\tilde{\Xi}_{12}$ &$\frac{1}{9009}{\rm T}^4{\rm h}$&
$-\frac{1}{28} {\rm T}^2$ \\ 
&  \\  
$\tilde{\Xi}_{13}$ &$\frac{1}{297}{\rm T}^2{\rm h}$&
$\frac{1}{12}$ \\ 
&  \\
$\tilde{\Xi}_{14}$ &$\frac{31}{3861}{\rm T}^2{\rm h}$&
$-\frac{1}{12}$ \\ 
&  \\ \\ \\
\hline  
\end{tabular}
\end{minipage}
\end{table*}

Squares of the dual functions $\tilde{\xi}_a$ appearing 
in eqs. (\ref{dz0})-(\ref{dz14}) and eqs. (\ref{dzT})-(\ref{dzTT}) is rather
complicated. Their behavior is periodic with bumps both near zero and orbital 
frequencies with rapidly decaying occilating wings far outside these 
frequencies. We evaluated that the frequency bump for the spin-down parameters 
near zero frequency 
is bigger than that near the orbital one. On the other hand, the situation for 
orbital parameters is just opposite. Analitic behavior of  
the dual functions
shows that there are two spectral windows in which they are the most sensitive
to the stochastic noise. One is located near zero frequency and the second
one lies near the orbital one. We can bound these windows by two
frequency intervals (0,$\frac{\alpha}{\rm T}$) and 
($1-\frac{\alpha_{-}}{\rm T},1+\frac{\alpha_{+}}{\rm T}$) respectively where 
constant
coefficients $\alpha$, $\alpha_{-}$, $\alpha_{+}$ can be calculated by means 
of comparison of results of calculation
of variances in time and frequency domains. It
allows easily to calculate contributions from the foregoing frequency bands to
numerical values of variances of fitting parameters and determine which of
these frequency bands makes bigger deposit. Variances of fitting parameters
depend only on the total span, $T$, of observations (Kopeikin 1997b). 
Comparing dependence of the variances on $T$ calculated in time domain with 
that calculated in two frequency windows reveal the relative importance of 
different frequencies.
In order to do this we have to compare magnitude of two integrals
\vspace{0.3 cm}
\begin{equation}
{\rm I_1} \sim h_n \displaystyle{\int_{\varepsilon}^{\frac{\alpha}{\rm T}}}
|\tilde{\Xi}_a(f)|^2 \left[\displaystyle{
\frac{1}{(2\pi f)^n}}+\displaystyle{\sum_{k=0}^\infty}B_{2k}(\varepsilon) 
\varepsilon^{2k}\delta^{(2k)}(f-\varepsilon)\right] df,
\label{i1}
\vspace{0.3 cm}
\end{equation}
and\vspace{0.3 cm}
\begin{equation}
{\rm I_2} \sim \frac{h_n}{(2\pi)^n}
\displaystyle{\int_{1-\frac{\alpha_{-}}{\rm T}}^
{1+\frac{\alpha_{+}}{\rm T}}}
|\tilde{\Xi}_a(f)|^2 f^{-n} df,
\label{i2}
\vspace{0.3 cm}
\end{equation} 

The first integral describes contribution of low frequencies to parameter's
variances. The second integral gives contribution to parameter's variances from 
the frequencies lying near the orbital one. It is worth emphasizing that terms 
with delta function and its derivatives in (\ref{i1}) bring on mutual 
cancellation of all terms depending on cut-off frequency $\varepsilon$ and 
diverging as $\varepsilon$ goes to zero. 
Such cancellation has been expected since we modified the spectrum of
red noise so that to avoid appearance of all divergent terms which have no 
physical meaning. We don't give here results of calculation of numerical values
of parameters $\alpha$, $\alpha_{-}$, $\alpha_{+}$ because they are not so
important for making conclusions. Asymptotic behavior of the dual functions
near zero and orbital frequency is enough to see which frequncy band is the
most important for giving contribution to corresponding integrals and 
parameter's variances. 
Time dependence of two integrals is shown in Table (\ref{tab12})
up to not so important constant.   
\begin{table*}
\begin{minipage}{130mm}
\centering
\caption{Comparative contribution of different frequency bands to variances of
spin-down ($a=1,2,...,6$) and orbital ($a=7,8,...,14$) parameters $\beta_a$.
Number $n=1,2,...,6$ denotes the spectral index of corresponding red noise.
Time dependence of all variances completely coincides with that which was
obtained by calculations in time domain as given in (Kopeikin 1999).}
\label{tab12}
\vspace{3 mm}
\begin{tabular}{|ccc|}
\hline \\ 
Variance of parameter&Contribution of integral ${\rm I_1}$&
Contribution of integral ${\rm I_2}$ \\ \hline
&  \\ 
$\sigma_{\beta_1}^2$&$\sim {\rm T}^{n-1}$&$\sim {\rm T}^{-3}$ \\ 
&  \\ 
$\sigma_{\beta_2}^2$&$\sim {\rm T}^{n-3}$&$\sim {\rm T}^{-5}$ \\ 
&  \\ 
$\sigma_{\beta_3}^2$&$\sim {\rm T}^{n-5}$&$\sim {\rm T}^{-7}$ \\ 
&  \\ 
$\sigma_{\beta_4}^2$&$\sim {\rm T}^{n-7}$&$\sim {\rm T}^{-9}$ \\ 
&  \\ 
$\sigma_{\beta_5}^2$&$\sim {\rm T}^{n-9}$&$\sim {\rm T}^{-11} $\\ 
&  \\ 
$\sigma_{\beta_6}^2$&$\sim {\rm T}^{n-11}$&$\sim {\rm T}^{-13}$ \\ 
&  \\
$\sigma_{\beta_7}^2$&$\sim {\rm T}^{n-5}$&$\sim {\rm T}^{-1}$ \\ 
&  \\ 
$\sigma_{\beta_8}^2$&$\sim {\rm T}^{n-3}$&$\sim {\rm T}^{-1}$ \\ 
&  \\ 
$\sigma_{\beta_9}^2$&$\sim {\rm T}^{n-5}$&$\sim {\rm T}^{-3}$ \\ 
&  \\  
$\sigma_{\beta_{10}}^2$&$\sim {\rm T}^{n-7}$&$\sim {\rm T}^{-3}$ \\ 
&  \\ 
$\sigma_{\beta_{11}}^2$&$\sim {\rm T}^{n-9}$&$\sim {\rm T}^{-5}$ \\ 
&  \\
$\sigma_{\beta_{12}}^2$&$\sim {\rm T}^{n-7}$&$\sim {\rm T}^{-5}$ \\ 
&  \\  
$\sigma_{\beta_{13}}^2$&$\sim {\rm T}^{n-9}$&$\sim {\rm T}^{-7}$ \\ 
&  \\
$\sigma_{\beta_{14}}^2$&$\sim {\rm T}^{n-11}$&$\sim {\rm T}^{-7}$ \\ 
&  \\ \\
\hline  
\end{tabular}
\end{minipage}
\end{table*}
Behavior of variances of the first three spin-down parameters is not
interesting for they are contaminated by the presence of the non-stationary 
part of red noise (Kopeikin 1997b). For the rest of spin-down parameters one 
observes that contribution of noise energy from  low frequencies to variances 
of the parameters is dominating. However, the situation is not so simple in 
the event of orbital parameters. 
One can see that in the event of red noise having spectral index $n \leq 4$
contribution of the noise energy from the orbital frequency span 
($1-\frac{\alpha_{-}}{\rm T},1+\frac{\alpha_{+}}{\rm T}$) can be equal to or
even bigger than that from the low frequency band. Only when the spectral index 
of noise $n\geq 5$ contribution of the noise energy of low frequencies to 
variances of orbital parameters begins to dominate. It is worth noting that the
timing noise with spectral index $n=5$ is produced by cosmological
gravitational wave background. The fact that for this noise 
low-frequencies give the main contribution to variances of orbital parameters
confirms our early statement (Kopeikin 1997a) that measurement of variances of
orbital parameters showing secular evolution tests the ultra-low frequency band
of cosmological gravitational wave background. Hence, these variances can be 
used for setting upper limit on the cosmological parameter $\Omega_g$ in this
frequency range in contrast to timing residuals which test only low-frequency
band of the background noise.  
(Kopeikin 1997a).  
\section{Acknowledgments}
 
We are grateful to N. Wex
for numerous fruitful discussions which have helped to improve the 
presentation of the manuscript. 
%We thank D. Moran
%for careful reading of the manuscript and valuable comments. 
S.M. Kopeikin is
pleasured to acknowledge the hospitality of G. Neugebauer and G. Sch\"afer and
other members of the Institute for Theoretical Physics of the Friedrich
Schiller University of Jena. 
This work has been partially
supported by the Th\"uringer Ministerium f\"ur Wissenschaft, Forschung und 
Kultur grant No B501-96060.

%\newpage

\appendix

\section{Explicit expressions for the dual functions}

In this appendix we give explicit expressions for the dual functions. Using
definition (\ref{gfl}) and elements of inverse matrix $C_{ab}^{-1}$ from the
paper (Kopeikin 1999, Tables 5 and 6) one obtains
\vspace{0.3 cm}
\begin{eqnarray}\label{dfu1}
\tilde{\Xi}_1(z)&=&j_0(z)+\frac{5}{2}j_2(z)+\frac{27}{8}j_4(z)+
\\\nonumber\\\mbox{}&&
+\frac{15{\rm h}}{8\rm T}\biggl\{3\left[j_1(z+{\rm T})-j_1(z-{\rm T})\right]
-7\left[j_3(z+{\rm T})-j_3(z-{\rm T})\right]\biggr\}\\\nonumber\\\mbox{}&&-
\frac{45{\rm h}}{8{\rm T}^2}\biggl\{3\left[j_0(z+{\rm T})+ j_0(z-{\rm T})\right]
-20\left[j_2(z+{\rm T})+j_2(z-{\rm T})\right]\biggr\}\;,
\end{eqnarray}

\begin{eqnarray}\label{dfu2}
\frac{1}{i}\tilde{\Xi}_2(z)&=&-\frac{3}{\rm T}\left[
j_1(z)+\frac{7}{2}j_3(z)+\frac{55}{8}j_5(z)\right]+
\\\nonumber\\\mbox{}&&
\frac{105{\rm h}}{8{\rm T}^2}
\biggl\{j_0(z+{\rm T})-j_0(z-{\rm T})
-5\left[j_2(z+{\rm T})-j_2(z-{\rm T})\right]\biggr\}+\\\nonumber\\\mbox{}&&
\frac{105{\rm h}}{8{\rm T}^3}\biggl\{51\left[
j_1(z+{\rm T})+ j_1(z-{\rm T})\right]
-154\left[j_3(z+{\rm T})+j_3(z-{\rm T})\right]\biggr\}\;,
\end{eqnarray}
 
\begin{eqnarray}\label{dfu3}
\tilde{\Xi}_3(z)&=&
-\frac{15}{2{\rm T}^2}\left[j_2(z)+\frac{9}{2}j_4(z)\right]
\\\nonumber\\\mbox{}&&-
\frac{105{\rm h}}{4{\rm T}^3}\biggl\{3\left[j_1(z+{\rm T})-j_1(z-{\rm T})\right]
-7\left[j_3(z+{\rm T})-j_3(z-{\rm T})\right]\biggr\}+\\\nonumber\\\mbox{}&&
\frac{105{\rm h}}{4{\rm T}^4}\biggl\{7\left[j_0(z+{\rm T})+ j_0(z-{\rm T})\right]
-50\left[j_2(z+{\rm T})+j_2(z-{\rm T})\right]\biggr\}\;,
\end{eqnarray} 

\begin{eqnarray}\label{dfu4}
\frac{1}{i}\tilde{\Xi}_4(z)&=&\frac{35}{2{\rm T}^3}\left[
j_3(z)+\frac{11}{2}j_5(z)\right]
\\\nonumber\\\mbox{}&&-
\frac{315{\rm h}}{4{\rm T}^4}\biggl\{j_0(z+{\rm T})-j_0(z-{\rm T})
-5\left[j_2(z+{\rm T})-j_2(z-{\rm T})\right]\biggr\}\\\nonumber\\\mbox{}&&-
\frac{1575{\rm h}}{4{\rm T}^5}\biggl\{9\left[
j_1(z+{\rm T})+ j_1(z-{\rm T})\right]
-28\left[j_3(z+{\rm T})+j_3(z-{\rm T})\right]\biggr\}\;,
\end{eqnarray}

\begin{eqnarray}\label{dfu5}
\tilde{\Xi}_5(z)&=&\frac{315}{8{\rm T}^4}\;j_4(z)
\\\nonumber\\\mbox{}&&+
\frac{315{\rm h}}{8{\rm T}^5}\biggl\{3\left[j_1(z+{\rm T})-j_1(z-{\rm T})\right]
-7\left[j_3(z+{\rm T})-j_3(z-{\rm T})\right]\biggr\}\\\nonumber\\\mbox{}&&-
\frac{1575{\rm h}}{8{\rm T}^6}\biggl\{\left[j_0(z+{\rm T})+ j_0(z-{\rm T})\right]
-8\left[j_2(z+{\rm T})+j_2(z-{\rm T})\right]\biggr\}\;,
\end{eqnarray}

\begin{eqnarray}\label{dfu6}
\frac{1}{i}\tilde{\Xi}_6(z)&=&-\frac{693}{8{\rm T}^5}\;j_5(z)+
\\\nonumber\\\mbox{}&&
\frac{693{\rm h}}{8{\rm T}^6}\biggl\{j_0(z+{\rm T})-j_0(z-{\rm T})
-5\left[j_2(z+{\rm T})-j_2(z-{\rm T})\right]\biggr\}+\\\nonumber\\\mbox{}&&
\frac{2079{\rm h}}{8{\rm T}^7}\biggl\{13\left[
j_1(z+{\rm T})+ j_1(z-{\rm T})\right]
-42\left[j_3(z+{\rm T})+j_3(z-{\rm T})\right]\biggr\}\;,
\end{eqnarray}

\begin{eqnarray}\label{dfu7}
\tilde{\Xi}_7(z)&=&j_0(z+{\rm T})+ j_0(z-{\rm T})+\frac{5}{2}
\left[j_2(z+{\rm T})+j_2(z-{\rm T})\right]
\\\nonumber\\\mbox{}&&-
\frac{3}{4{\rm T}}\biggl\{3\left[j_1(z+{\rm T})-j_1(z-{\rm T})\right]
-7\left[j_3(z+{\rm T})-j_3(z-{\rm T})\right]\biggr\}\\\nonumber\\\mbox{}&&-
\frac{45{\rm h}}{{\rm T}^2}\left[j_2(z)-6j_4(z)\right]\;,
\end{eqnarray}

\begin{eqnarray}\label{dfu8}
\frac{1}{i}\tilde{\Xi}_8(z)&=&
j_0(z+{\rm T})-j_0(z-{\rm T})+
\frac{5}{2}\left[j_2(z+{\rm T})-j_2(z-{\rm T})\right]+
\\\nonumber\\\mbox{}&&
\frac{3}{4{\rm T}}\biggl\{3\left[j_1(z+{\rm T})+j_1(z-{\rm T})\right]
-7\left[j_3(z+{\rm T})+j_3(z-{\rm T})\right]\biggr\}+
\\\nonumber\\\mbox{}&&
\frac{3{\rm h}}{{\rm T}}
\left[3j_1(z)-7j_3(z)+11j_5(z)\right]\;,
\end{eqnarray}

\begin{eqnarray}\label{dfu9}
\tilde{\Xi}_9(z)&=&\frac{3}{{\rm T}}\biggl\{j_1(z+{\rm T})-j_1(z-{\rm T})+
\frac{7}{2}\left[j_3(z+{\rm T})-j_3(z-{\rm T})\right]\biggr\}
\\\nonumber\\\mbox{}&&-
\frac{15}{4{\rm T}^2}\biggl\{\left[j_0(z+{\rm T})+ j_0(z-{\rm T})\right]
-5\left[j_2(z+{\rm T})+j_2(z-{\rm T})\right]\biggr\}
\\\nonumber\\\mbox{}&&-
\frac{15{\rm h}}{{\rm T}^2}\left[j_0(z)-5j_2(z)+9j_4(z)\right]\;,
\end{eqnarray}

\begin{eqnarray}\label{dfu10}
\frac{1}{i}\tilde{\Xi}_{10}(z)&=&
-\frac{3}{{\rm T}}\biggl\{\left[j_1(z+{\rm T})+j_1(z-{\rm T})\right]+
\frac{7}{2}\left[j_3(z+{\rm T})+j_3(z-{\rm T})\right]\biggr\}
\\\nonumber\\\mbox{}&&-
\frac{15}{4{\rm T}^2}\biggl\{j_0(z+{\rm T})-j_0(z-{\rm T})-
5\left[j_2(z+{\rm T})-j_2(z-{\rm T})\right]+
\\\nonumber\\\mbox{}&&
\frac{15{\rm h}}{{\rm T}^3}
\left[-18j_1(z)+77j_3(z)-220j_5(z)\right]\;,
\end{eqnarray}

\begin{eqnarray}\label{dfu11}
\tilde{\Xi}_{11}(z)&=&
-\frac{15}{2{\rm T}^2}\left[j_2(z+{\rm T})+ j_2(z-{\rm T})\right]
\\\nonumber\\\mbox{}&&+
\frac{15}{4{\rm T}^3}\biggl\{3\left[j_1(z+{\rm T})-j_1(z-{\rm T})\right]
-7\left[j_3(z+{\rm T})-j_3(z-{\rm T})\right]\biggr\}+\\\nonumber\\\mbox{}&&
\frac{15{\rm h}}{{\rm T}^4}
\left[2j_0(z)+5j_2(z)-72j_4(z)\right]\;,
\end{eqnarray}

\begin{eqnarray}\label{dfu12}
\frac{1}{i}\tilde{\Xi}_{12}(z)&=&
-\frac{15}{2{\rm T}^2}\left[j_2(z+{\rm T})- j_2(z-{\rm T})\right]
\\\nonumber\\\mbox{}&&-
\frac{15}{4{\rm T}^3}\biggl\{3\left[j_1(z+{\rm T})+j_1(z-{\rm T})\right]
-7\left[j_3(z+{\rm T})+j_3(z-{\rm T})\right]\biggr\}+\\\nonumber\\\mbox{}&&
-\frac{15{\rm h}}{{\rm T}^3}
\left[3j_1(z)-7j_3(z)+11j_5(z)\right]\;,
\end{eqnarray}

\begin{eqnarray}\label{dfu13}
\tilde{\Xi}_{13}(z)&=&-
\frac{35}{2{\rm T}^3}\left[j_3(z+{\rm T})- j_3(z-{\rm T})\right]+
\\\nonumber\\\mbox{}&&
\frac{35}{4{\rm T}^4}\biggl\{j_0(z+{\rm T})+j_0(z-{\rm T})
-5\left[j_2(z+{\rm T})+j_2(z-{\rm T})\right]\biggr\}+\\\nonumber\\\mbox{}&&
\frac{35{\rm h}}{{\rm T}^4}
\left[j_0(z)-5j_2(z)+9j_4(z)\right]\;,
\end{eqnarray}

\begin{eqnarray}\label{dfu14}
\frac{1}{i}\tilde{\Xi}_{14}(z)&=&
\frac{35}{2{\rm T}^3}\left[j_3(z+{\rm T})+ j_3(z-{\rm T})\right]+
\\\nonumber\\\mbox{}&&
\frac{35}{4{\rm T}^4}\biggl\{j_0(z+{\rm T})-j_0(z-{\rm T})
-5\left[j_2(z+{\rm T})-j_2(z-{\rm T})\right]\biggr\}+\\\nonumber\\\mbox{}&&
\frac{105{\rm h}}{{\rm T}^5}
\left[4j_1(z)-21j_3(z)+66j_5(z)\right]\;.
\end{eqnarray}

\section{Asymptotic behavior of the dual functions near zero frequency}

In this appendix we give asymptotic behavior of the dual functions near zero
frequency. They are as follows:

\begin{eqnarray}\label{dz0}
\tilde{\Xi}_1(f)&=&\tilde{\xi}_1(z)\;,\quad\quad\tilde{\xi}_1(z)=
j_0(z)+\frac{5}{2}j_2(z)+\frac{27}{8}j_4(z)\;,\\\nonumber\\
\frac{1}{i}\tilde{\Xi}_2(f)&=&-\frac{3}{\rm T}\tilde{\xi}_2(z)\;,\quad\quad
\tilde{\xi}_2(z)=j_1(z)+\frac{7}{2}j_3(z)+\frac{55}{8}j_5(z)\;,
\\\nonumber\\
\tilde{\Xi}_3(f)&=&-\frac{15}{2{\rm T}^2}\tilde{\xi}_3(z)\;,\quad\quad
\tilde{\xi}_3(z)=j_2(z)+\frac{9}{2}j_4(z)\;,
\\\nonumber\\
\frac{1}{i}\tilde{\Xi}_4(f)&=&\frac{35}{2{\rm T}^3}\tilde{\xi}_4(z)\;,
\quad\quad\tilde{\xi}_4(z)=j_3(z)+\frac{11}{2}j_5(z)\;,
\\\nonumber\\
\tilde{\Xi}_5(f)&=&\frac{315}{8{\rm T}^4}\tilde{\xi}_5(z)\;,\quad\quad
\tilde{\xi}_5(z)=j_4(z)\;,\\\nonumber\\
\frac{1}{i}\tilde{\Xi}_6(f)&=&-\frac{693}{8{\rm T}^5}\tilde{\xi}_6(z)\;,
\quad\quad\tilde{\xi}_6(z)=j_5(z)\;,\\\nonumber\\
\tilde{\Xi}_7(f)&=&-\frac{45{\rm h}}{{\rm T}^2}\tilde{\xi}_7(z)\;,\quad\quad
\tilde{\xi}_7(z)=j_2(z)-6j_4(z)-\frac{1}{15}z\sin z\;,\\\nonumber\\
\frac{1}{i}\tilde{\Xi}_8(f)&=&\frac{9{\rm h}}{{\rm T}}\tilde{\xi}_8(z)\;,
\quad\quad\tilde{\xi}_8(z)=j_1(z)-\frac{7}{3}j_3(z)+\frac{11}{3}j_5(z)-
\frac{1}{3}\sin z\;,
\\\nonumber\\
\tilde{\Xi}_9(f)&=&-\frac{15{\rm h}}{{\rm T}^2}\tilde{\xi}_9(z)\;,\quad\quad
\tilde{\xi}_9(z)=j_0(z)-5j_2(z)+9j_4(z)-\cos z\;,
\\\nonumber\\
\frac{1}{i}\tilde{\Xi}_{10}(f)&=&-\frac{270{\rm h}}{{\rm T}^3}\tilde{\xi}_{10}(z)\;,
\quad\quad\tilde{\xi}_{10}(z)=j_1(z)-\frac{77}{18}j_3(z)+\frac{110}{9}j_5(z)-
\frac{5}{18}\sin z-\frac{1}{18}z \cos z\;,\\\nonumber\\
\tilde{\Xi}_{11}(f)&=&\frac{30{\rm h}}{{\rm T}^4}\tilde{\xi}_{11}(z)\;,\quad\quad
\tilde{\xi}_{11}(z)=j_0(z)+\frac{5}{2}j_2(z)-36j_4(z)-\frac{1}{2}z \sin z- 
\cos z\;,\\\nonumber\\
\frac{1}{i}\tilde{\Xi}_{12}(f)&=&-\frac{45{\rm h}}{{\rm T}^3}\tilde{\xi}_{12}(z)\;,
\quad\quad\tilde{\xi}_{12}(z)=j_1(z)-\frac{7}{3}j_3(z)+\frac{11}{3}j_5(z)-
\frac{1}{3}\sin z\;,
\\\nonumber\\
\tilde{\Xi}_{13}(f)&=&\frac{35{\rm h}}{{\rm T}^4}\tilde{\xi}_{13}(z)\;,
\quad\quad\tilde{\xi}_{13}(z)=\tilde{\xi}_{9}(z)\;,
\\\nonumber\\\label{dz14}
\frac{1}{i}\tilde{\Xi}_{14}(f)&=&\frac{420{\rm h}}{{\rm
T}^5}\tilde{\xi}_{14}(z)\;,\quad\quad\tilde{\xi}_{14}(z)=
j_1(z)-\frac{63}{12}j_3(z)+\frac{33}{2}j_5(z)-\frac{1}{4}\sin z-
\frac{1}{12}z\cos z\;,\\\nonumber
\\\nonumber
\end{eqnarray}
where ${\rm h}=(-1)^N$.

\section{Asymptotic behavior of the dual functions near orbital frequency}

Leading terms in asymptotic expansions of the dual functions $\tilde{\Xi}_a(z)$
near orbital frequency are as follows:

\begin{eqnarray}\label{dzT}
\tilde{\Xi}_1(f)&=&-\frac{45\rm h}{8{\rm T}}\tilde{\xi}_1(y)\;,\quad\quad
\tilde{\xi}_{1}(y)=j_1(y)-\frac{7}{3}j_3(y)-\frac{1}{3}\sin y\;,\\\nonumber
\\
\frac{1}{i}\tilde{\Xi}_2(f)&=&-\frac{105\rm h}{8{\rm T}^2}\tilde{\xi}_2(y)\;,\quad\quad
\tilde{\xi}_2(y)=j_0(y)-5j_2(y)-\cos y\;,\\\nonumber\\
\tilde{\Xi}_3(f)&=&-\frac{14}{{\rm T}^3}\tilde{\xi}_3(y)\;,\quad\quad
\tilde{\xi}_3(y)=\tilde{\xi}_{1}(y)\\\nonumber
\\
\frac{1}{i}\tilde{\Xi}_4(f)&=&-\frac{6}{{\rm T}^4}\tilde{\xi}_4(y)\;,\quad\quad
\tilde{\xi}_4(y)=\tilde{\xi}_2(y)\;,\\\nonumber \\
\tilde{\Xi}_5(f)&=&\frac{21}{{\rm T}^5}\tilde{\xi}_5(y)\;,\quad\quad
\tilde{\xi}_5(y)=\tilde{\xi}_{1}(y)\;,\\\nonumber\\
\frac{1}{i}\tilde{\Xi}_6(f)&=&\frac{33}{5{\rm T}^6}\tilde{\xi}_6(y)\;,\quad\quad
\tilde{\xi}_6(y)=\tilde{\xi}_2(y)\;,\\\nonumber\\
\tilde{\Xi}_7(f)&=&\tilde{\xi}_7(y)\;,\quad\quad
\tilde{\xi}_7(y)=j_0(y)+\frac{5}{2}j_2(y)\;,\\\nonumber\\
\frac{1}{i}\tilde{\Xi}_8(f)&=&\tilde{\xi}_8(y)\;,\quad\quad
\tilde{\xi}_8(y)=-\tilde{\xi}_7(y)\;,\\\nonumber\\
\tilde{\Xi}_9(f)&=&-\frac{3}{{\rm T}}\tilde{\xi}_9(y)\;,\quad\quad
\tilde{\xi}_9(y)=j_1(y)+\frac{7}{2}j_3(y)\;,\\\nonumber\\
\frac{1}{i}\tilde{\Xi}_{10}(f)&=&\frac{1}{{\rm T}}\tilde{\xi}_{10}(y)\;,\quad\quad
\tilde{\xi}_{10}(y)=\tilde{\xi}_9(y)\;,\\\nonumber\\
\tilde{\Xi}_{11}(f)&=&-\frac{15}{2{\rm T}^2}\tilde{\xi}_{11}(y)\;,\quad\quad
\tilde{\xi}_{11}(y)=j_2(y)\;,\\\nonumber\\
\frac{1}{i}\tilde{\Xi}_{12}(f)&=&\frac{1}{{\rm T}^2}\tilde{\xi}_{12}(y)\;,\quad\quad
\tilde{\xi}_{12}(y)=-\tilde{\xi}_{11}(y)\;,\\\nonumber\\
\tilde{\Xi}_{13}(f)&=&\frac{35}{2{\rm T}^3}\tilde{\xi}_{13}(y)\;,\quad\quad
\tilde{\xi}_{13}(y)=j_3(y)\;,\\\nonumber\\\label{dzTT}
\frac{1}{i}\tilde{\Xi}_{14}(f)&=&\frac{1}{{\rm T}^3}\tilde{\xi}_{14}(y)\;,\quad\quad
\tilde{\xi}_{14}(y)=\tilde{\xi}_{13}(y)\;.\\\nonumber
\end{eqnarray}
where $y=z-{\rm T}$.

\section{Plots of the Fourier transform of spherical Bessel functions}

\newpage
\section{Plots of the Fourier transform of autocovariance function}
\clearpage
\begin{figure*}
\centerline{\psfig{figure=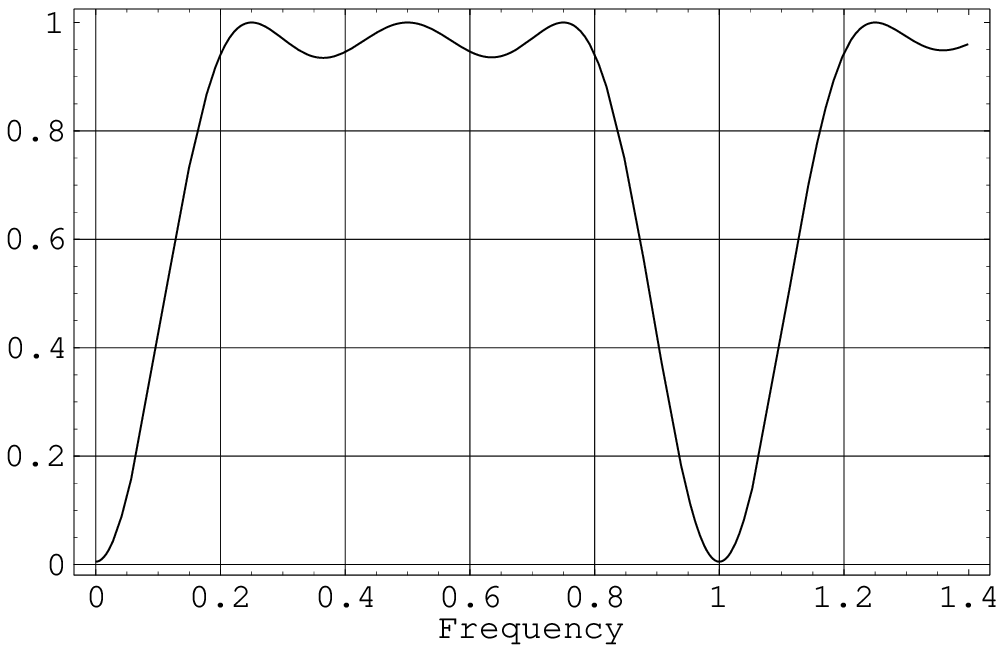,angle=0,height=8cm,width=17cm}}
\caption{Plot of the Fourier transform of  the filter function of 
timing residuals for the amount of
orbital revolutions $N=4$. Frequency is
measured in units of orbital frequency $n_b$. Amplitude of the transform has
been normalized to unity.}
\label{figa1}
\end{figure*}
\begin{figure*}
\centerline{\psfig{figure=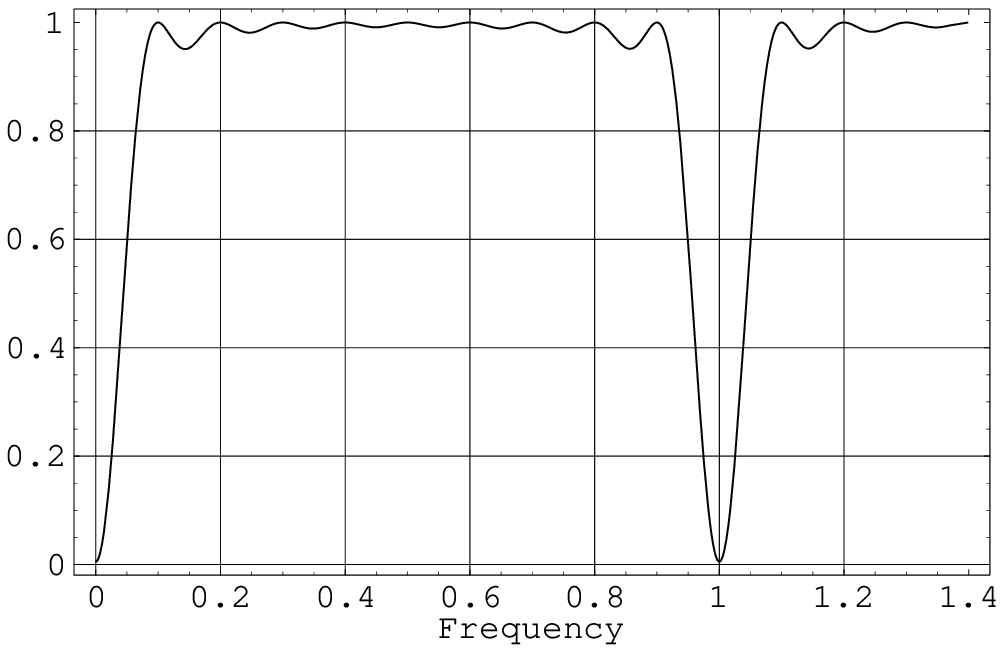,angle=0,height=8cm,width=17cm}}
\caption{Plot of the Fourier transform of  the filter function of 
timing residuals for the amount of
orbital revolutions $N=10$. Frequency is
measured in units of orbital frequency $n_b$. Amplitude of the transform has
been normalized to unity.}
\label{figb1}
\end{figure*}
\begin{figure*}
\centerline{\psfig{figure=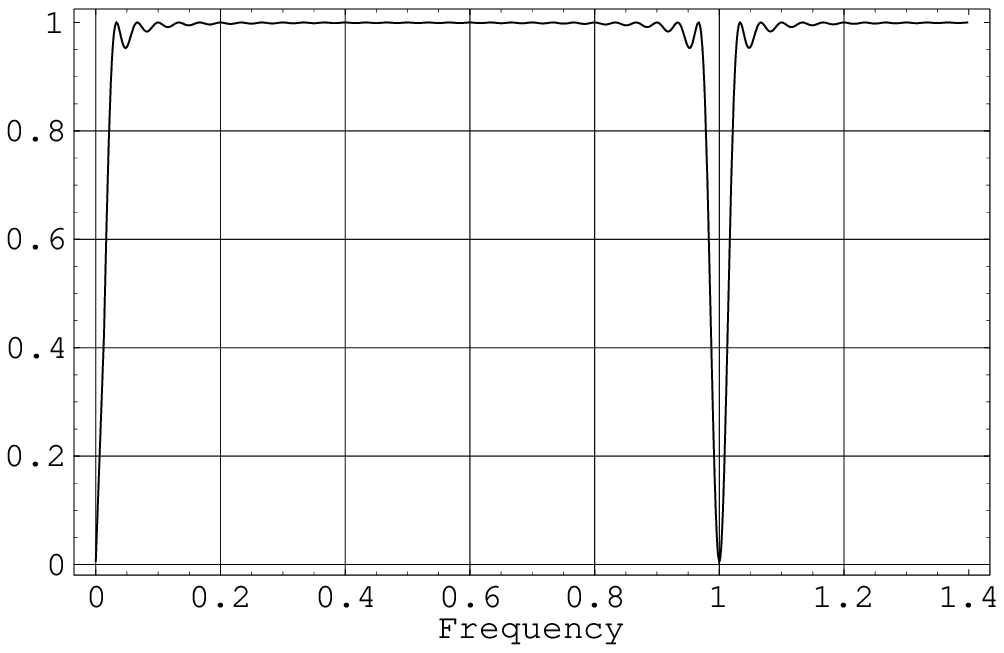,angle=0,height=8cm,width=17cm}}
\caption{Plot of the Fourier transform of the filter function of 
timing residuals for the amount of
orbital revolutions $N=30$. Frequency is
measured in units of orbital frequency $n_b$. Amplitude of the transform has
been normalized to unity.}
\label{figc1}
\end{figure*}

\label{lastpage}

\end{document}